\documentclass[aps]{revtex4-1}
\usepackage{amsfonts}
\usepackage{amssymb}
\usepackage{amsmath}
\usepackage{amsthm}
\usepackage{bm}
\usepackage{braket}
\usepackage{graphicx}
\usepackage{dcolumn}

\begin{document}

\title{A solvable model for symmetry-breaking phase transitions}
\author{Shatrughna Kumar$^{1}$}
\author{Pengfei Li$^{2}$}
\author{Liangwei Zeng$^{3}$}
\author{Jingsong He$^{4}$}
\author{Boris A. Malomed$^{*}$$^{1,5}$}
\address{$^{*}$The corresponding author\\
$^{1}$Department of Physical Electronics, School of Electrical Engineering, Faculty of Engineering, and Center for
Light-Matter Interaction, Tel Aviv University, Tel Aviv 69978, Israel\\
$^{2}$Department of Physics, Taiyuan Normal University, Jinzhong
030619, China\\
$^{3}$Department of Basic Course, Guangzhou Maritime University,
Guangzhou 510725, China\\
$^{4}$Institute for Advanced Study, Shenzhen University, Shenzhen,
Guangdong, China\\
$^{5}$Instituto de Alta Investigaci\'{o}n,
Universidad de Tarapac\'{a}, Casilla 7D, Arica, Chile}

\begin{abstract}
\textbf{Abstract}

Analytically solvable models are benchmarks in studies of phase transitions
and pattern-forming bifurcations. Such models are known for phase
transitions of the second kind in uniform media, but not for localized
states (solitons), as integrable equations which produce solitons do not
admit intrinsic transitions in them. We introduce a solvable model for
symmetry-breaking phase transitions of both the first and second kinds
(alias sub- and supercritical bifurcations) for solitons pinned to a
combined linear-nonlinear double-well potential, represented by a symmetric
pair of delta-functions. Both self-focusing and defocusing signs of the
nonlinearity are considered. In the former case, exact solutions are
produced for symmetric and asymmetric solitons. The solutions explicitly
demonstrate a switch between the symmetry-breaking transitions of the first
and second kinds (i.e., sub- and supercritical bifurcations, respectively).
In the self-defocusing model, the solution demonstrates the transition of
the second kind which breaks antisymmetry of the first excited state.
\end{abstract}

\maketitle

\section*{Introduction}

\subsection*{The topic: spontaneous symmetry breaking in nonlinear systems}

Dynamics of collective excitations in physical systems is determined by the
interplay of the underlying diffraction or dispersion, nonlinear
self-interactions of the fields or wave functions, and potentials acting on
the fields. In this context, it is commonly known that the ground state (GS)
of linear systems reproduces the symmetry of the underlying potential, while
excited states may realize other representations of the same symmetry \cite%
{LL}. In particular, the wave function of a particle trapped in a symmetric
double-well potential (DWP) is even, while the first excited state is odd.

While these basic properties are demonstrated by the linear Schr\"{o}dinger
equation, the dynamics of Bose-Einstein condensates (BECs) is governed, in
mean-field approximation, by the Gross-Pitaevskii equation (GPE), which
takes into regard interactions between particles, adding the cubic term to
the Schr\"{o}dinger equation for the single-particle wave function \cite%
{BEC1,BEC2}. The repulsive or attractive interactions are represented by the
cubic term with the self-defocusing (SDF) or self-focusing (SF) sign.
Essentially the same model is the celebrated nonlinear Schr\"{o}dinger
equation (NLSE), which governs the propagation of optical waves in nonlinear
media \cite{NLS} and finds other realizations, as the universal model to
govern the interplay of the weak diffraction or dispersion and cubic SF
nonlinearity \cite{Peyrard}. In optics, a counterpart of the trapping
potential is the term in the NLSE which accounts for the waveguiding
structure induced by a transverse profile of the refractive index.

The GS\ structure in models combining the DWP and SF nonlinearity follows
the symmetry of the underlying potential only in the weakly nonlinear
regime. A generic effect which occurs with the increase of the nonlinearity
strength is the symmetry-breaking phase transition, which makes the GS
asymmetric with respect to two wells of the DWP \cite{book}. This effect of
the spontaneous symmetry breaking (SSB) implies, \textit{inter alia}, that
the commonly known principle of quantum mechanics, according to which GS
cannot be degenerate \cite{LL}, is no longer valid in the nonlinear models:
obviously, the SSB gives rise to a degenerate pair of two mutually symmetric
GSs, with the maximum of the wave function pinned to the left or right
potential well of the underlying DWP. The same system admits a symmetric
state coexisting with the asymmetric ones, but, above the SSB point, it does
not represents the GS, being unstable against symmetry-breaking
perturbations.

In systems with the SDF sign of the nonlinearity, the GS remains symmetric
and stable, while the SSB transition breaks the \textit{antisymmetry} of the
first excited state (it is a spatially odd one, with precisely one zero of
the wave function, located at the central point). The resulting state with
the spontaneously broken antisymmetry keeps the zero point, which is shifted
from the center to right or left.

The concept of the SSB in systems of the NLSE type with the SF nonlinearity
was first proposed, in an abstract mathematical form, by Davies in 1979 \cite%
{Davies}. Another early realization of this concept was introduced in 1985
by Eilbeck, Lomdahl, and Scott, in the form of the \textquotedblleft
self-trapping model" \cite{Scott}. The latter work had actually initiated
systematic studies of SSB phase transitions.

In optics, the SSB was observed experimentally in a photorefractive crystal
with saturable SF nonlinearity and an effective DWP waveguiding structure
\cite{photo}. SSB was also predicted for photonic modes supported by a
symmetrically designed plasmonic metamaterial \cite{nano}. For the
self-attractive BEC loaded into a DWP trap, the symmetry-breaking transition
was elaborated in Refs. \cite{Milburn,Smerzi}. In that context,
tunnel-coupling oscillations between condensates trapped in two potential
wells separated by a barrier represent the bosonic Josephson effect \cite%
{junction}. Experimentally, the self-trapping of a stationary state with
spontaneously broken antisymmetry in a self-repulsive condensate loaded in
DWP, as well as Josephson oscillations in that setup, were reported in Ref.
\cite{Markus}.

A ramification of the topic is SSB in dual-core systems, such as twin-core
optical fibers, with the SF cubic nonlinearity acting in each core \cite%
{Jensen}. In such systems, the interplay of the SF and linear coupling
between the parallel cores gives rise to the SSB transition from the
symmetric state to a spontaneously established one with unequal powers
carried by the two cores. This type of the symmetry-breaking phase
transition was studied in detail theoretically \cite{Wabnitz}-\cite{Pak} and
recently demonstrated experimentally in a twin-core fiber \cite{Bugar}. In
terms of the respective system of linearly-coupled NLSEs, the SSB transition
is represented by the bifurcation which links symmetric and asymmetric
solutions \cite{bif}. Depending on the type of the intra-core nonlinearity
and the wave form under the consideration (delocalized or self-trapped), the
symmetry-breaking bifurcation may be of the supercritical (alias forward) or
subcritical (backward) type. The corresponding bifurcations give rise to the
destabilization of the symmetric state and creation of the pair of
asymmetric ones at the SSB point, which go forward or backward as stable or
unstable branches, respectively, following the variation of the SSB-driving
nonlinearity strength. In the latter (subcritical) case, the unstable
\textit{lower} branches of the asymmetric states normally reverse into the
stable forward-going \textit{upper} ones at certain turning points (see,
e.g., Fig. \ref{fig5} below). As a result, stable asymmetric states emerge
subcritically, at a value of the nonlinearity strength which is smaller than
that at the SSB point. Accordingly, the system is bistable in the interval
between the turning and SSB points, where the symmetric and upper asymmetric
states coexist as stable ones. In terms of statistical physics, the super-
and subcritical bifurcations are identified as symmetry-breaking phase
transitions of the second and first kinds, respectively. In the latter case,
the bistability corresponds to the hysteresis between the GS and overcooled
or overheated phases.

Other varieties of optical SSB effects occur in dual-core laser setups
combining the SF nonlinearity with gain and loss. The theoretical model of
such setups is based on a pair of linearly coupled complex Ginzburg-Landau
equations with the cubic-quintic nonlinearity \cite{Sigler}. A spontaneously
established asymmetric regime of the operation of a symmetric pair of
coupled lasers was observed in Ref. \cite{lasers}.

The SSB phenomenology was also predicted in models with symmetric \textit{%
nonlinear potentials}, induced by spatial modulation of the local SF or SDF
coefficient \cite{Barcelona}. In optical waveguides, the modulation can be
imposed by spatially inhomogeneous distributions of a resonant dopant, which
gives rise to strong local nonlinearity \cite{Kip}. In experiments with BEC,
the Feshbach resonance (FR) controlled by spatially nonuniform laser
illumination of the condensate may be employed to build an effective
nonlinearity landscape \cite{Bauer,Yamazaki,Clark}. Other techniques
available to the experimental work with BEC\ make it possible to
\textquotedblleft paint" a necessary FR-induced nonlinear potential by a
fast moving laser beam \cite{Henderson} or a spatial light modulator \cite%
{Hadzibabic,Nogrette,Bruce}.

\subsection*{The model}

The use of the nonlinear potential suggests a possibility to design
experimentally feasible \textit{solvable} SSB\ settings, which admit exact
analytical solutions for symmetric, antisymmetric, and asymmetric states.
The key component of solvable models is the nonlinear term in the NLSE with
coordinate $x$, which is concentrated at $x=0$, being represented by the $%
\delta $-function:%
\begin{equation}
i\frac{\partial \psi }{\partial z}=-\frac{1}{2}\frac{\partial ^{2}\psi }{%
\partial x^{2}}-\delta \left( x\right) \left( \varepsilon +\sigma |\psi
|^{2}\right) \psi .  \label{single-delta}
\end{equation}%
This model is formulated in terms of optics, with the evolution along
propagation distance $z$ under the action of the real nonlinearity
coefficient $\sigma $, scaled to be $\sigma =+1$ or $-1$, which corresponds,
respectively, to the SF or SDF sign of the nonlinearity. In that case, the $%
\delta $-function term represents a narrow layer of an optical material with
strong cubic susceptibility (e.g., AlGaAs, whose susceptibility exceeds that
of silica by a factor $\simeq 700$ \cite{AlGaAs}) embedded in the linear
planar waveguide, provided that the width of the layer is small in
comparison with that of self-trapped light beams propagating in the
waveguide. This setting can be readily implemented in the experiment, as the
typical width of spatial solitons is measured in tens of microns \cite%
{Silberberg}. In that case, the linear trapping potential, $-\varepsilon
\delta (x)$, present in Eq. (\ref{single-delta}), is relevant too, as the
linear refractive index of materials such as AlGaAs is much higher than the
background value in the host material (silica). As concerns the sign of the
nonlinearity, the consideration of the SDF layer is also interesting, as
semiconductor materials may demonstrate negative nonlinear susceptibility.

The same equation (\ref{single-delta}), with $z$ replaced by time $t$,
applies to BEC, with the $\delta $-function potential induced by the
FR-inducing laser beam tightly focused at $x=0$. The same optical beam also
induces the linear potential represented by coefficient $\varepsilon $. In a
similar context, Eq. (\ref{single-delta}) with $\varepsilon =0$ was first
introduced, as a model of a nonlinear bosonic junction, in Ref. \cite{Azbel}%
. Further, a model of the matter-wave soliton interferometer with a
nonlinear soliton splitter corresponds to $\varepsilon <0$ and $\sigma =-1$
in Eq. (\ref{single-delta}) \cite{NJP}.

Equation (\ref{single-delta}) gives rise to the exact solution for a family
of self-trapped states (solitons) pinned to the delta-functional potential:%
\begin{equation}
\psi _{0}(x,z)=U_{0}(x)\exp (ikz),  \label{EU}
\end{equation}%
where $k$ is an arbitrary propagation constant, and the shape function is%
\begin{equation}
U_{0}(x)=\sqrt{\sigma \left( \sqrt{2k}-\varepsilon \right) }\exp \left( -%
\sqrt{2k}|x|\right) .  \label{U}
\end{equation}%
The self-trapped modes are characterized by their integral power,%
\begin{equation}
P_{0}=\int_{-\infty }^{+\infty }U_{0}^{2}(x)dx=\sigma \left( 1-\frac{%
\varepsilon }{\sqrt{2k}}\right) ,  \label{P}
\end{equation}%
which is a dynamical invariant of Eq. (\ref{single-delta}).

The well-known Vakhitov-Kolokolov (VK) criterion, $dP/dk>0$ \cite%
{VK,Berge,Sulem,Fibich}, immediately implies that the family of solutions (%
\ref{U}) in the case of the SF nonlinearity, $\sigma =+1$, and $\varepsilon
>0$ is stable in its entire existence region, $k>\varepsilon ^{2}/2$ (and
completely unstable if the linear potential is repulsive, with $\varepsilon
<0$). For localized states supported by the SDF nonlinearity, with $\sigma
=-1$, the VK\ stability criterion is replaced by the anti-VK one \cite{HS}, $%
dP/dk<0$. Accordingly, in this case the localized states (\ref{U}) are also
stable in their entire existence region, which is $0<k<\varepsilon ^{2}/2$.
The definition of the power given by Eq. (\ref{P}) demonstrates that the
bound states pinned to $\delta $-function potential with the SF sign of the
nonlinearity exist in the interval of $0<P_{0}<1$, while the competition of
the linear attractive potential and SDF nonlinear term gives rise to the
bound states in the entire range of $0<P_{0}<\infty $.

An exceptional case is the one corresponding to $\sigma =+1$ (SF) and $%
\varepsilon =0$ (no linear potential), for which Eq. (\ref{P}) demonstrates
the degeneracy of the localized states, whose power takes the single value, $%
P_{0}=1$, which does not depend on $k$. This property implies that the
corresponding family represents a specific example of Townes solitons (a
commonly known family of Townes solitons is one produced by localized
solutions of two-dimensional NLSE with the spatially uniform cubic SF
nonlinearity \cite{Townes}). Because Townes solitons, with their single
value of the power, have $dP/dk=0$, the VK criterion predicts that they
correspond to a border between the stability and instability. It is known
that, in fact, the Townes solitons are subject to the subexponentially
commencing instability, which eventually leads to the onset of the critical
collapse (emergence of a local singularity after a finite propagation
distance) \cite{Sulem,Fibich}.

It is also worthy to mention the value of the Hamiltonian of the pinned
state (\ref{U}),%
\begin{eqnarray}
H_{0} &=&\frac{1}{2}\int_{-\infty }^{+\infty }\left\vert \frac{\partial \psi
_{0}}{\partial x}\right\vert ^{2}dx-\left[ \varepsilon \left\vert \psi
_{0}\left( x=0\right) \right\vert ^{2}+\frac{\sigma }{2}\left\vert \psi
_{0}\left( x=0\right) \right\vert ^{4}\right]  \notag \\
&=&-\frac{\sigma \varepsilon }{2}\left( \sqrt{2k}-\varepsilon \right)
\label{H0}
\end{eqnarray}%
[the Hamiltonian is another dynamical invariant of Eq. (\ref{single-delta}%
)]. Note that the existence condition for solution (\ref{U}), $\sigma \left(
\sqrt{2k}-\varepsilon \right) >0$, implies $H_{0}<0$ for $\varepsilon >0$,
hence the localized solution represents a true bound state with the negative
energy.

The possibility to produce exact analytical solutions for localized states
pinned to the $\delta $-function nonlinear potential suggests a possibility
to design a solvable DWP model based on a set of two $\delta $-functions,
separated by distance which may be set equal to $1$ by means of rescaling:%
\begin{equation}
i\frac{\partial \psi }{\partial z}=-\frac{1}{2}\frac{\partial ^{2}\psi }{%
\partial x^{2}}-\left[ \delta \left( x+\frac{1}{2}\right) +\delta \left( x-%
\frac{1}{2}\right) \right] \left( \varepsilon +\sigma |\psi |^{2}\right)
\psi .  \label{psi}
\end{equation}%
The equation for stationary states is produced by the substitution of
expression (\ref{EU}) in Eq. (\ref{psi}):%
\begin{equation}
kU-\frac{1}{2}\frac{d^{2}U}{dx^{2}}-\left[ \delta \left( x+\frac{1}{2}%
\right) +\delta \left( x-\frac{1}{2}\right) \right] \left( \varepsilon
+\sigma U^{2}\right) U=0.  \label{stationary}
\end{equation}%
The Hamiltonian corresponding to Eq. (\ref{psi}) is%
\begin{equation}
H=\frac{1}{2}\int_{-\infty }^{+\infty }\left\vert \frac{\partial \psi }{%
\partial x}\right\vert ^{2}dx-\sum_{\pm }\left[ \varepsilon \left\vert \psi
\left( x=\pm \frac{1}{2}\right) \right\vert ^{2}+\frac{\sigma }{2}\left\vert
\psi \left( x=\pm \frac{1}{2}\right) \right\vert ^{4}\right] ,
\label{Hamiltonian}
\end{equation}%
cf. Eq. (\ref{H0}). The physical implementation of the model in optics and
BEC is straightforward: in the former case, one can embed two parallel
nonlinear layers in the linear waveguide, while in the former case the
necessary configuration may be created by two tightly focused FR-inducing
laser beams.

A particular case of Eq. (\ref{psi}) with $\varepsilon =0$ was introduced,
in the context of BEC, in Ref. \cite{Thawatchai}. Exact solutions for
symmetric, antisymmetric, and, which is most interesting, asymmetric
stationary wave functions were produced in that work, demonstrating a very
peculiar feature, namely, an SSB bifurcation of the \textit{extreme
subcritical type}, in which backward-going branches of unstable states never
turn forward and, accordingly, never become stable. In other words, it is a
unique example of the symmetry-breaking phase transition of the first kind
which does not produce any stable phase past the transition point, and gives
rise to a fully unstable overcooled phase, represented by the completely
unstable asymmetric states.

Recently, another example of such an anomalous phase transition was found in
Ref. \cite{Strunin} in the study of dual-core couplers with the SF
nonlinearity and fractional diffraction, represented by operator $\left(
-\partial ^{2}/\partial x^{2}\right) ^{\alpha /2}$, with \textit{L\'{e}vy
index} $\alpha $ \cite{Levy}, acting in each core. In that case, the extreme
subcritical SSB\ bifurcation takes place at $\alpha =1$, which is the border
between the normal symmetry-breaking phase transition of the first kind at $%
1<\alpha <2$, and full instability of the system, driven by the
supercritical collapse, at $\alpha <1$. However, the fractional-coupler
model cannot be solved analytically, on the contrary to Eq. (\ref{stationary}%
).

\subsection*{Objectives of the work}

Our purpose is to produce an analytical solution of the full model, with the
combined linear-nonlinear $\delta $-functional DWP in Eq. (\ref{psi}). The
linear terms are represented by $\varepsilon >0$ (the attractive potential),
while both SF and SDF signs of the nonlinearity, $\sigma =\pm 1$, will be
addressed. For $\sigma =+1$, the solution explicitly demonstrates gradual
switch from the extreme subcritical bifurcation to the supercritical one via
a regular subcritical bifurcation, in which the backward-going (lower)
branches of unstable asymmetric states reverse into stable upper branches at
turning points. For $\sigma =-1$ the results are more straightforward,
corroborating the stability of the symmetric GS and the occurrence of the
supercritical antisymmetry-breaking transition in the first excited state.

While the asset of the model is its analytical solvability, some results are
produced in a numerical form, using Eqs. (\ref{psi}) and (\ref{stationary})
with a regularized $\delta $-function,%
\begin{equation}
\tilde{\delta}(x)=\left( \pi w\right) ^{-1/2}\exp \left( -x^{2}/w^{2}\right)
,  \label{tilde}
\end{equation}%
defined by a small width $w$ (in most cases, $w=0.01$ was used, which is $%
1/100$ of the distance between the two $\delta $-funtions). In this
connection, it is relevant to mention that the realization of the present
model as the optical waveguide implies that a characteristic value of the
separation between the two narrow attractive layers may be $\sim 50$ $%
\mathrm{\mu }$m, hence $w=0.01$ corresponds to the layer's thickness $\sim
0.5$ $\mathrm{\mu }$m. In view of the above-mentioned possibility to use a
material with the nonlinear susceptibility exceeding that in the bulk
waveguide by a factor $\sim 700$, this thickness will be sufficient to
provide the requires nonlinearity. In the case of the realization in BEC, a
relevant size of the separation may be $\sim 10$ $\mathrm{\mu }$m. Then, the
nearly-delta-functional potential may be induced by a laser beam focused on
a spot of size $\sim $ $0.5$ $\mathrm{\mu }$m, which will correspond to $%
w\simeq 0.05$, in the scaled units.

The comparison with the numerical solutions is relevant to check how well
the solvable model represents a realistic one, with the finite width $w$ of
the potential wells, and also to test predictions for stability of the
symmetric, antisymmetric, and asymmetric solitons pinned to the $\delta $%
-functional DWP. The analytical and numerical results are summarized in the
next section, and are discussed in the concluding one.

\section*{Results}

\subsection*{Exact analytical solutions for the symmetric, asymmetric, and
antisymmetric states}

\subsubsection*{The self-focusing nonlinearity}

The fact that Eq. (\ref{stationary}) is linear at $x\neq \pm 1/2$ makes it
possible to construct obvious solutions in these areas, as $\exp \left( -%
\sqrt{2k}\left\vert x+1/2\right\vert \right) $ and $\exp \left( -\sqrt{2k}%
\left\vert x-1/2\right\vert \right) $ at $x<-1/2$ and $x>+1/2$,
respectively, and a combination of these terms at $|x|<1/2$. At points $%
x=\pm 1/2$, the solutions are matched by the continuity condition for $U(x)$
and the jump condition for the derivative $dU/dx$,
\begin{equation}
dU/dx|_{x\pm 1/2=+0}-dU/dx|_{x\pm 1/2=-0}=-2\left( \varepsilon +\sigma
U^{2}\right) U|_{x\pm 1/2=0}~.  \label{jump}
\end{equation}%
The generic solution satisfying these conditions can be looked for as
\begin{eqnarray}
U(x) &=&U_{1}(k)\exp \left( \sqrt{2k}\left( x+\frac{1}{2}\right) \right) ,~%
\mathrm{at}~x<-\frac{1}{2},  \label{left} \\
U(x) &=&U_{2}(k)\exp \left( \sqrt{2k}\left( -\left\vert x-\frac{1}{2}%
\right\vert \right) \right) ,~\mathrm{at}~x>\frac{1}{2},  \label{right} \\
U(x) &=&U_{2}(k)\frac{\sinh \left( \sqrt{2k}\left( x+1/2\right) \right) }{%
\sinh \left( \sqrt{2k}\right) }+U_{1}(k)\frac{\sinh \left( \sqrt{2k}\left(
1/2-x\right) \right) }{\sinh \left( \sqrt{2k}\right) },\mathrm{at}~|x|<\frac{%
1}{2},  \label{middle}
\end{eqnarray}%
where amplitudes $U_{1}(k)$ and $U_{2}(k)$ should be found from the
substitution of ansatz (\ref{left})-(\ref{middle}) in Eq. (\ref{jump}). For
this stationary solution, the value of Hamiltonian (\ref{Hamiltonian}) is%
\begin{equation}
H=2kP-\varepsilon \left( U_{1}^{2}+U_{2}^{2}\right) -\frac{\sigma }{2}\left(
U_{1}^{4}+U_{2}^{4}\right) ,  \label{H(N)}
\end{equation}%
where $P$ is the integral power, defined as per Eq. (\ref{P}).

\begin{figure}[tbph]
\includegraphics[scale=0.25]{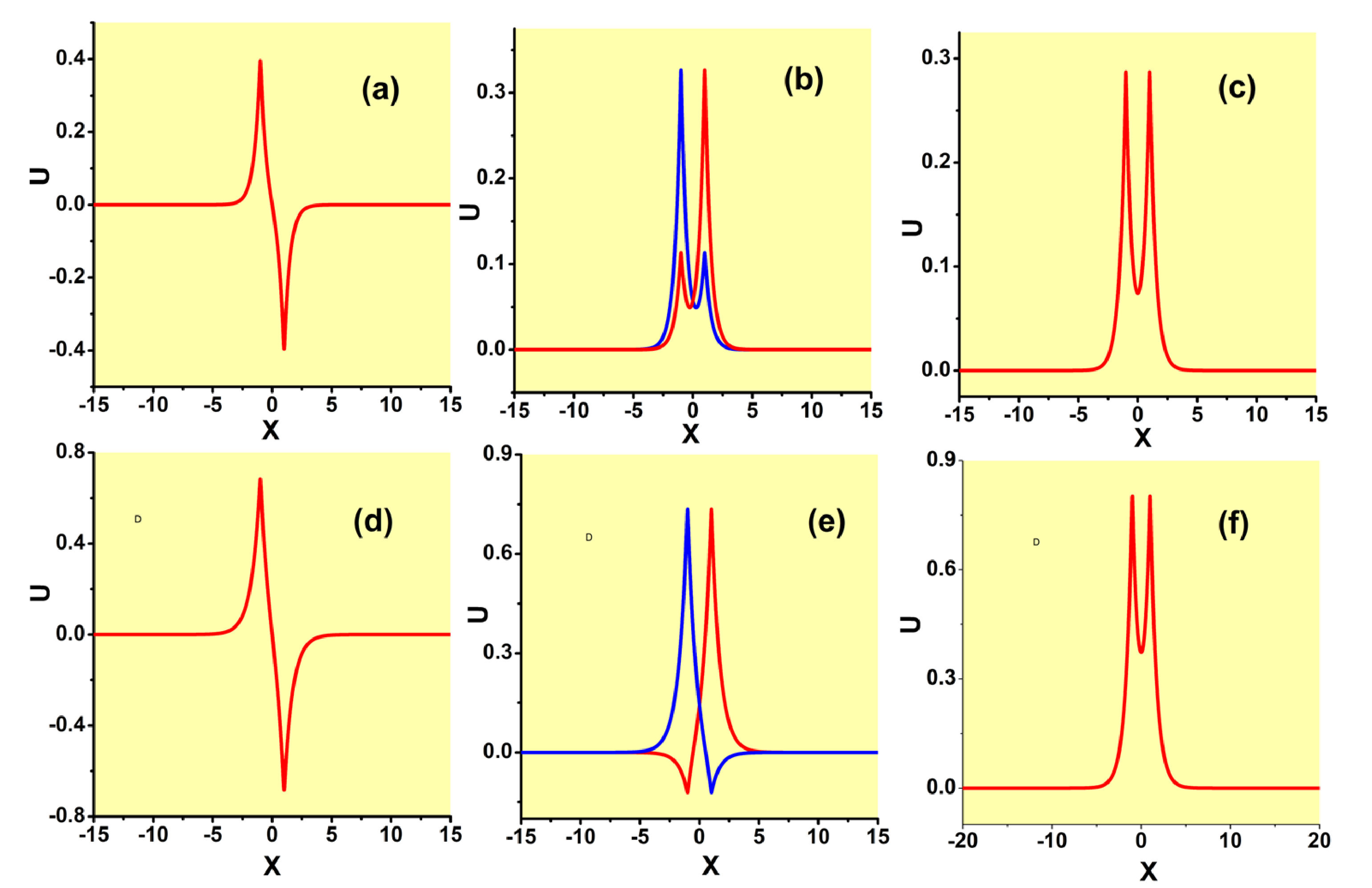}
\caption{(a-c) Typical examples of antisymmetric, asymmetric, and symmeric
bound states (solitons) produced by the numerical solution of Eq. (\protect
\ref{stationary}) with the $\protect\delta $-functions approximated by
expression (\protect\ref{tilde}), the SF nonlinearity ($\protect\sigma =+1$%
), and parameters $\protect\varepsilon =2$, $k=2.1$. In panel (b) two
asymmetric states are plotted, which are mirror images of each other. The
antisymmetric and symmetric states are unstable, while the asymmetric one is
stable. (d-f) Typical examples of antisymmetric, broken-antisymmetry, and
symmetric bound states for the SDF nonlinearity ($\protect\sigma =-1$) and $%
\protect\varepsilon =2$, $k=1$. In panel (e), two states with broken
antisymmetry are mirror images of each other. The antisymmetric state is
unstable, while the ones with broken antisymmetry and unbroken symmetry are
stable.}
\label{fig0}
\end{figure}

First, it is easy to find the exact solutions for symmetric states in the
SF\ model ($\sigma =+1$), with equal amplitudes $U_{1}(k)=U_{2}(k)\equiv U_{%
\mathrm{symm}}(k)$:%
\begin{equation}
U_{\mathrm{symm}}(k)=\sqrt{\sigma \left[ E(\varepsilon ,k)-1\right] /S(k)},
\label{Usymm}
\end{equation}%
where

\begin{eqnarray}
S(k) &\equiv &\sqrt{\frac{2}{k}}\sinh \left( \sqrt{2k}\right) ,  \label{S} \\
E(\varepsilon ,k) &\equiv &\exp \left( \sqrt{2k}\right) -\varepsilon \sqrt{%
\frac{2}{k}}\sinh \left( \sqrt{2k}\right) .  \label{F}
\end{eqnarray}%
A typical example of a symmetric bound state (soliton), for $\varepsilon =2$
and $k=2.1$, is displayed in Fig. \ref{fig0}(c). This plots is produced by
the numerical solution of Eq. (\ref{stationary}), being virtually
indistinguishable from its counterpart given by the analytical solution, as
provided by Eqs. (\ref{left})-(\ref{middle}) and (\ref{Usymm}).

\begin{figure}[tbph]
\includegraphics[scale=0.04]{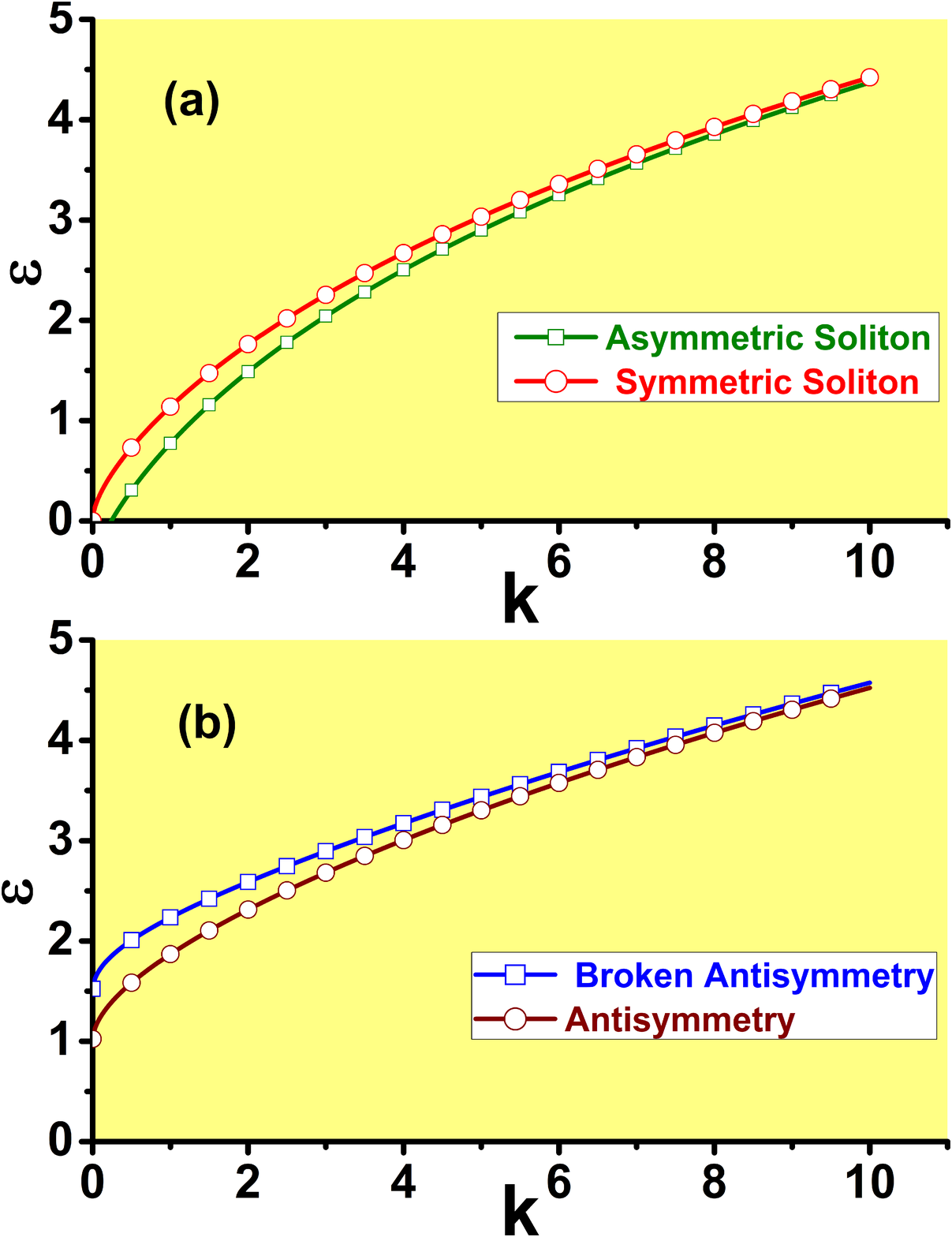}
\caption{(a) In the model with the SF nonlinearity, $\protect\sigma =+1$,
the symmetric bound states with amplitudes (\protect\ref{Usymm}) exist
beneath the boundary in the plane of $\left( k,\protect\varepsilon \right) $
displayed by the red curve, which is produced by Eq. (\protect\ref{max-symm}%
). The asymmetric states, with the amplitudes given by Eqs. (\protect\ref{U1}%
) and (\protect\ref{U2}), exist beneath the green boundary, which is
produced by Eq. (\protect\ref{crit}). (b) In the model with the SDF
nonlinearity, $\protect\sigma =-1$, the antisymmetric bound states with
amplitudes (\protect\ref{Usymm}) exist above the brown boundary, which is
defined by Eq. (\protect\ref{max-symm}). The states with broken antisymmetry
and amplitudes given by Eqs. (\protect\ref{U1anti}) and (\protect\ref{U2anti}%
) exist above the blue boundary, which is defined by Eq. (\protect\ref%
{crit-anti}).}
\label{fig1}
\end{figure}

Because $S$, as defined by Eq. (\ref{S}), is always positive, the solution
given by Eq. (\ref{Usymm}) for $\sigma =+1$ and $-1$ exists for $%
E(\varepsilon ,k)>1$ and $E(\varepsilon ,k)<1$, respectively. As it follows
from Eq. (\ref{F}), this condition implies that, in the case of SF
nonlinearity, the symmetric state with given propagation constant $k$ exists
if the strength of the linear $\delta $-function potential does not exceed a
maximum value,%
\begin{equation}
\varepsilon <\left( \varepsilon _{\max }\right) _{\mathrm{symm}}\equiv \frac{%
\sqrt{2k}}{1+\exp \left( -\sqrt{2k}\right) }.  \label{max-symm}
\end{equation}%
In other words, for given $\varepsilon $ the symmetric state exists for $k$
exceeding a value $\left( k_{\min }\right) _{\mathrm{symm}}$ determined by
Eq. (\ref{max-symm}) with\ $<$ replaced by $=$, i.e., beneath the red curve
in Fig. \ref{fig1}(a). In particular,
\begin{equation}
\left( k_{\min }\right) _{\mathrm{symm}}\approx \left\{
\begin{array}{c}
2\varepsilon ^{2},~\mathrm{for}~\varepsilon \ll 1, \\
\varepsilon ^{2}/2,~\mathrm{for}~\varepsilon \gg 1.%
\end{array}%
\right.  \label{kmin}
\end{equation}%
In the SDF case, the existence area for the symmetric states is opposite, $%
\varepsilon >\left( \varepsilon _{\max }\right) _{\mathrm{symm}}$. The
existence boundary (\ref{max-symm}) is shown by the red curve in Fig. \ref%
{fig1}(a).

\begin{figure}[tbph]
\includegraphics[scale=0.04]{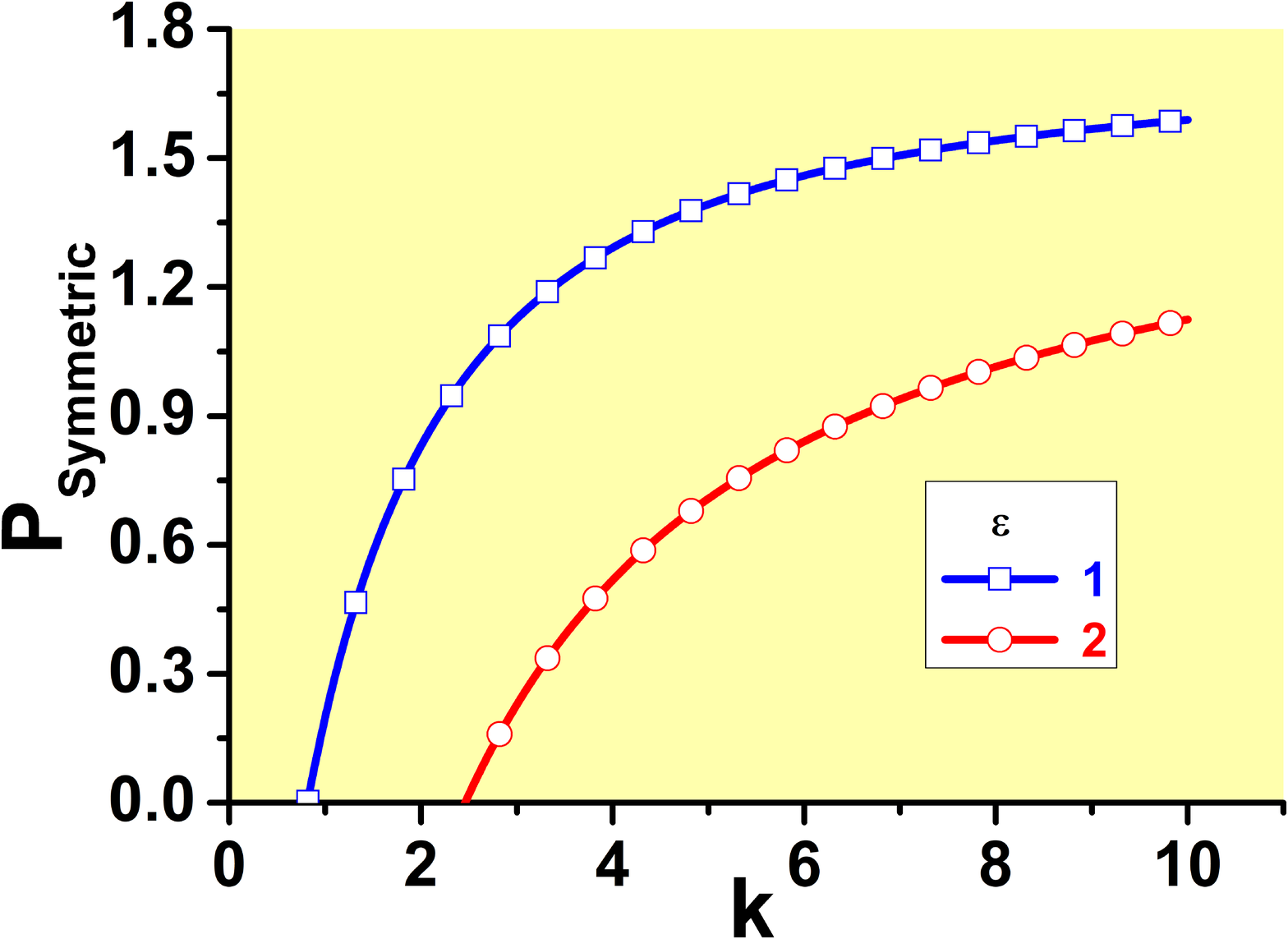}
\caption{The dependence of the integral power of the symmetric bound states
on the propagation constant, in the model with the SF sign of the
nonlinearity, as given by Eq. (\protect\ref{Psymm}), for $\protect%
\varepsilon =1$ and $2$. As shown by Eq. (\protect\ref{P_symm-max}), with
the increase of $k$ the power is slowly approaching the limit value, $P_{%
\mathrm{symm}}(k=\infty )=2$.}
\label{fig2}
\end{figure}

The bound states (solitons) are characterized by the total power defined as
per Eq. (\ref{P}). For the symmetric states in the model with the SF
nonlinearity, it is%
\begin{equation}
P_{\mathrm{symm}}(k)\equiv \int_{-\infty }^{+\infty }U^{2}(x)dx=\frac{1}{2}%
U_{\mathrm{symm}}^{2}(k)\left[ \frac{1+\tanh \left( \sqrt{k/2}\right) }{%
\sqrt{k/2}}+\frac{1}{\cosh ^{2}\left( \sqrt{k/2}\right) }\right] .
\label{Psymm}
\end{equation}%
As $k$ varies from the minimum value $\left( k_{\min }\right) _{\mathrm{symm}%
}$ [see Eq. (\ref{kmin})] towards $k\rightarrow \infty $, the power (\ref%
{Psymm}) grows from $0$ to $2$, so that
\begin{equation}
P_{\mathrm{symm}}(k\rightarrow \infty )=2\left( 1-\varepsilon /\sqrt{2k}%
\right) .  \label{P_symm-max}
\end{equation}%
Examples of this dependence for $\varepsilon =1$ and $2$ are displayed in
Fig. \ref{fig2}. Note that it satisfies the above-mentioned VK criterion, $%
dP/dk>0$.

An essential fact is that the substitution of ansatz (\ref{left})-(\ref%
{middle}) in Eq. (\ref{jump}) produces, as well, an \emph{exact solution}
for asymmetric bound states in the model with the SF nonlinearity, with the
following values of amplitudes $U_{1}$ and $U_{2}$:

\begin{eqnarray}
\left( U_{\mathrm{asy}}\right) _{1}(k) &=&\sqrt{\frac{E(k)+\sqrt{E^{2}(k)-4}%
}{2S(k)}},  \label{U1} \\
\left( U_{\mathrm{asy}}\right) _{2}(k) &=&\sqrt{\frac{E(k)-\sqrt{E^{2}(k)-4}%
}{2S(k)}}  \label{U2}
\end{eqnarray}%
(or $U_{1}\rightleftarrows U_{2}$). Typical examples of stable asymmetric
states are presented in Fig. \ref{fig0}(b). They are produced as numerical
solutions of Eq. (\ref{stationary}), being indistinguishable from their
analytically found counterparts.

Obviously, the solution given by Eqs. (\ref{U1}) and (\ref{U2}) bifurcates
from the symmetric one (\ref{Usymm}) (with $\sigma =+1$) at $E=2$, and
exists at $E>2$. For a given propagation constant, the asymmetric solution
exists if $\varepsilon $ does not exceed a respective maximum value,
\begin{equation}
\varepsilon <\left( \varepsilon _{\max }\right) _{\mathrm{asy}}\equiv \sqrt{%
2k}\frac{1-2\exp \left( -\sqrt{2k}\right) }{1-\exp \left( -2\sqrt{2k}\right)
}<\left( \varepsilon _{\max }\right) _{\mathrm{symm}},  \label{crit}
\end{equation}%
cf. Eq. (\ref{max-symm}). The boundary (\ref{crit}) is shown by the green
curve in Fig. \ref{fig1}(a). For fixed $\varepsilon $, the asymmetric
solution exists in the region beneath this boundary, and only the symmetric
state exists in the stripe between the red and green curves in Fig. \ref%
{fig1}(a). In particular, $\left( \varepsilon _{\max }\right) _{\mathrm{symm}%
}(k\rightarrow 0)=0$, i.e., at $\varepsilon =0$ the symmetric states exist
in the entire region of $0<k<\infty $, while it follows from Eq. (\ref{crit}%
) that, in the same limit of $\varepsilon \rightarrow 0$, the asymmetric
state exists, in agreement with Ref. \cite{Thawatchai}, at
\begin{equation}
k>\left( k_{\min }\right) _{\mathrm{asy}}\left( \varepsilon =0\right) \equiv
(1/2)\left( \ln 2\right) ^{2}\approx 0.24.  \label{eps=0}
\end{equation}%
In accordance with generic properties of the SSB bifurcation \cite{bif}, the
symmetric states are stable solely in the stripe between the red and green
curves in Fig. \ref{fig1}(a), being destabilized by the SSB bifurcation
beneath the green one. These expectations are corroborated below by direct
simulations of the perturbed evolution of the symmetric modes displayed in
Fig. \ref{fig10}.

The asymmetry degree of stationary states is defined, in terms of the
respective integral power,
\begin{equation}
P\left( k\right) =\int_{-\infty }^{0}U^{2}(x)dx+\int_{0}^{+\infty
}U^{2}(x)dx\equiv P_{-}+P_{+}~,  \label{NN}
\end{equation}%
as%
\begin{equation}
\Theta \equiv \frac{P_{+}-P_{-}}{P_{+}+P_{-}}.  \label{Theta}
\end{equation}%
Full analytical expressions for the integral power of the asymmetric states,
$P_{\mathrm{asy}}(k)$, and the respective value of $\Theta $ are very
cumbersome. Nevertheless, it is easy to find that, while $k$ grows from the
minimum value $\left( k_{\min }\right) _{\mathrm{asy}}(\varepsilon )$ at the
SSB bifurcation point, which is determined by the left inequality in Eq. (%
\ref{crit}) replaced by the equality [see, in particular, Eq. (\ref{eps=0})
for $\varepsilon =0$], towards $k\rightarrow \infty $, $P_{\mathrm{asy}}(k)$
varies from the bifurcation-point value,%
\begin{equation}
P_{\mathrm{bif}}=P_{\mathrm{symm}}\left( k=\left( k_{\min }\right) _{\mathrm{%
asy}}(\varepsilon )\right) ,  \label{Pbif}
\end{equation}%
[with $P_{\mathrm{symm}}(k)$ given by Eq. (\ref{Psymm})] to
\begin{equation}
P_{\mathrm{asy}}(k\rightarrow \infty )=1.  \label{1}
\end{equation}%
Actually, Eq. (\ref{1}) is the same value as given above by Eq. (\ref{P})
with $\sigma =+1$ and $k\rightarrow \infty $. It follows from the above
expressions that, as $\varepsilon $ increases from zero towards infinity, $%
P_{\mathrm{bif}}$ monotonously decreases from
\begin{equation}
P_{\mathrm{bif}}\left( \varepsilon =0\right) =\frac{8}{27}\left( 3+\ln
2\right) \approx 1.094  \label{Pbif-max}
\end{equation}%
to $P_{\mathrm{bif}}\left( \varepsilon \rightarrow \infty \right) =0$. In
particular, $P_{\mathrm{bif}}\left( \varepsilon \right) $ is exponentially
small for large $\varepsilon $:%
\begin{equation}
P_{\mathrm{bif}}\left( \varepsilon \right) \approx \exp \left( -\varepsilon
\right) .  \label{exp}
\end{equation}

\begin{figure}[tbph]
\includegraphics[scale=0.25]{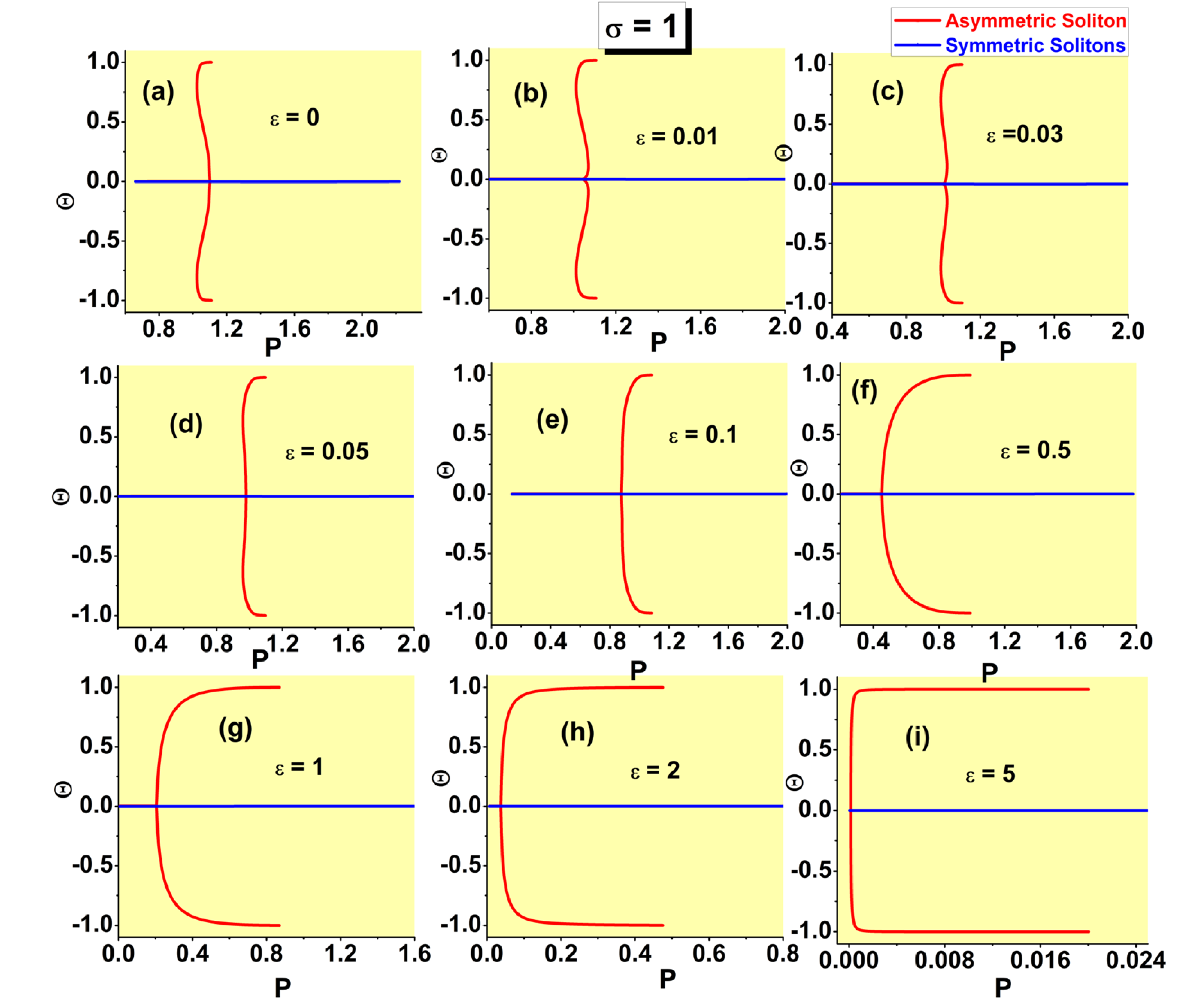}
\caption{The asymmetry parameter (\protect\ref{Theta}) for the numerically
produced solutions of Eq. (\protect\ref{stationary}) with the SF
nonlinearity ($\protect\sigma =+1$) vs. the integral power (\protect\ref{NN}%
) at different values of strength $\protect\varepsilon $ of the linear $%
\protect\delta $-functional potential, which are indicated in panels. The
switch between the symmetry-breaking phase transitions of the first and
second kinds, alias sub- and supercritical SSB bifurcations, takes place
between $\protect\varepsilon =0.07$ and $0.08$, in agreement with analytical
result (\protect\ref{thr}).}
\label{fig3}
\end{figure}

Comparison of limit values (\ref{Pbif}) and (\ref{1}) of the integral power
for the asymmetric states makes it possible to identify a threshold value $%
\varepsilon _{\mathrm{thr}}$ for the switch of the SSB phase transition
between the first and second kinds (i.e., the switch between the sub- and
supercritical SSB bifurcation): the phase transition may only be of the
first kind for $P_{\mathrm{bif}}>$ $P_{\mathrm{asy}}(k\rightarrow \infty
)\equiv 1$, and it becomes the second-order transition for $P_{\mathrm{bif}%
}<1$. The corresponding equation, $P_{\mathrm{bif}}=1$, combined with Eq. (%
\ref{crit}), in which $\varepsilon <\left( \varepsilon _{\max }\right) _{%
\mathrm{asy}}$ is replaced, as said above, by $\varepsilon =\left(
\varepsilon _{\max }\right) _{\mathrm{asy}}$, amounts to%
\begin{equation}
1+\tanh \left( \sqrt{k_{\mathrm{thr}}/2}\right) +\sqrt{k_{\mathrm{thr}}/2}%
\mathrm{sech}^{2}\left( \sqrt{k_{\mathrm{thr}}/2}\right) -2\sinh (\sqrt{2k_{%
\mathrm{thr}}})=0,  \label{eq}
\end{equation}%
where $k_{\mathrm{thr}}\equiv \left( k_{\min }\right) _{\mathrm{asy}}\left(
\varepsilon =\varepsilon _{\mathrm{thr}}\right) $. Numerical solution of Eq.
(\ref{eq}) produces the single root, $k_{\mathrm{thr}}\approx \allowbreak
0.298$, with the respective threshold value of $\varepsilon $ produced by
Eq. (\ref{crit}):%
\begin{equation}
\varepsilon _{\mathrm{thr}}\approx 0.074.  \label{thr}
\end{equation}%
This result is corroborated by comparison with numerically generated SSB
diagrams, in the form of $\Theta (P)$ dependences, which are displayed in
Fig. \ref{fig3}. In a detailed form, the numerical data demonstrate that the
threshold value belongs to interval $0.07<\varepsilon _{\mathrm{thr}}<0.08$,
while it is difficult to extract $\varepsilon _{\mathrm{thr}}$ from the data
with higher accuracy.

Note that narrow intervals of the variation of $P$ for branches of the
asymmetric states in panels (a-f) of Fig. \ref{fig3} correspond to the
analytical results presented here [see, e.g., the limits of the variation of
$P$ given by Eqs. (\ref{1}) and (\ref{Pbif-max})]; in panels (g-i), the $%
\Theta (P)$ curves are partly cut. The range of the variation of $P$ for the
branch of the symmetric states, with $\Theta \equiv 0$, is chiefly
determined by the limit value (\ref{P_symm-max}). For $\varepsilon =0$, the
detailed analysis, reported for this case in Ref. \cite{Thawatchai},
demonstrates, in agreement with Fig. \ref{fig3}(a), that the largest power
of the symmetric solitons is $\left( P_{\mathrm{symm}}\right) _{\max
}\approx 2.08$, which is attained at $k\approx 1.40$.

The asymmetric solitons are completely stable in the area $\varepsilon _{%
\mathrm{thr}}<\varepsilon <\left( \varepsilon _{\max }\right) _{\mathrm{asy}%
} $, as illustrated below by Fig. \ref{fig11}. At $\varepsilon <\varepsilon
_{\mathrm{thr}}$, the asymmetric solutions belonging to the lower branches
in Figs. \ref{fig3}(a-d), with $d\Theta /dP<0$, are unstable, while the
upper branches, with $d\Theta /dP>0$, are stable. Actually, the instability
intervals for the asymmetric solitons are very narrow.

In addition to the symmetric and asymmetric stationary states, Eqs. (\ref%
{psi}) and (\ref{psi}) with the SF sign of the nonlinearity, $\sigma =+1$,
give rise to antisymmetric ones, with $U(-x)=-U(x)$, see an example in Fig. %
\ref{fig0}(a). However, as well as in the case of $\varepsilon =0$ \cite%
{Thawatchai}, the antisymmetric states are completely unstable because, for
the same value of integral power $P$, they correspond to higher values of
Hamiltonian (\ref{Hamiltonian}) than the symmetric bound states. The
instability of the antisymmetric states is illustrated below by Fig. \ref%
{fig12}.

\subsubsection*{The self-defocusing nonlinearity}

\begin{figure}[tbph]
\includegraphics[scale=0.25]{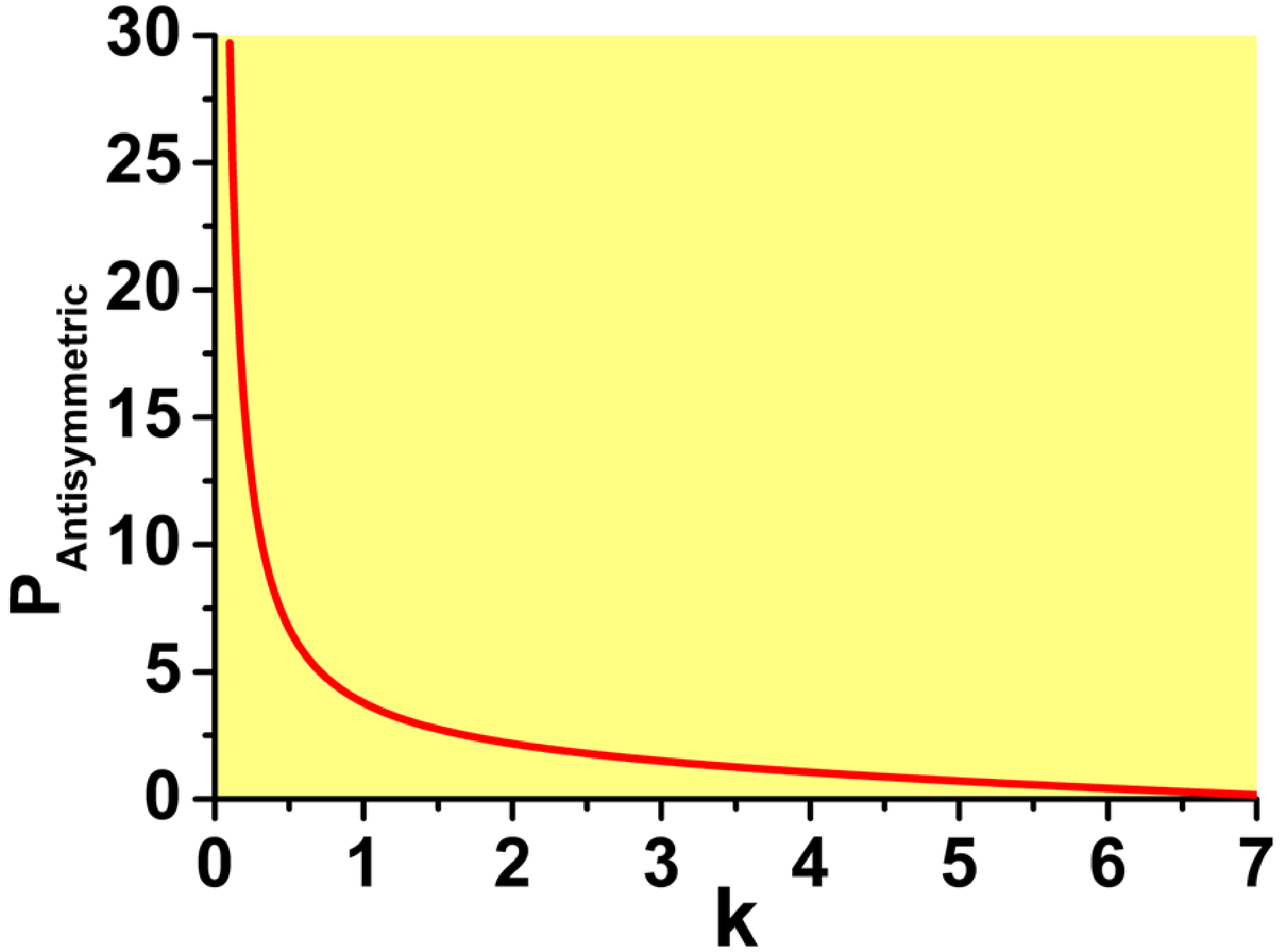}
\caption{The dependence of the integral power of the antisymmetric bound
states on the propagation constant $k$, in the case of the SDF sign of the
nonlinearity ($\protect\sigma =-1$), as given by Eq. (\protect\ref{Nantisymm}%
), for $\protect\varepsilon =4$. At $k\rightarrow 0$ the power diverges
according to Eq. (\protect\ref{divergence}). The power vanishes at $k\approx
7.686$, which is determined by Eq. (\protect\ref{anti}) with $\protect%
\varepsilon =4$.}
\label{fig2b}
\end{figure}

Typical examples of antisymmetric, broken-antisymmetry, and symmetric states
produced by Eq. (\ref{stationary}) with the SDF nonlinearity, i.e., $\sigma
=-1$ in (\ref{stationary}), are displayed in Figs. \ref{fig0}(d-f),
respectively. In this case, the symmetric state, given by solution (\ref%
{Usymm}) with $\sigma =-1$, which exists, as said above, at $\varepsilon
>\left( \varepsilon _{\max }\right) _{\mathrm{symm}}$ [see Eq. (\ref%
{max-symm})], is always stable, realizing the model's GS. Accordingly, it is
not subject to SSB. More interesting is the first excited state above the
GS, i.e., the antisymmetric one, given by the Eqs. (\ref{left})-(\ref{middle}%
) (with $\sigma =-1$)
\begin{equation}
U_{1}(k)=-U_{2}(k)=\sqrt{-\left[ E(k)+1\right] /S(k)}\equiv U_{\mathrm{anti}%
}(k),  \label{12anti}
\end{equation}%
where $S$ and $E$ is defined, as above, as per Eqs. (\ref{S}) and (\ref{F}).
Because $S$ is always positive, this solution exists under condition $E<-1$.
The substitution of Eq. (\ref{F}) demonstrates that this condition amounts to%
\begin{equation}
\varepsilon \geq \left( \varepsilon _{\min }\right) _{\mathrm{anti}}\equiv
\frac{\sqrt{2k}}{1-\exp \left( -\sqrt{2k}\right) },  \label{anti}
\end{equation}%
cf. Eq. (\ref{max-symm}).\ The antisymmetric state exists, at $\varepsilon >1
$, in the area of the $\left( \mu ,\varepsilon \right) $ plane above the
brown boundary shown in Fig. \ref{fig1}(b). Because Eq. (\ref{anti}) yields $%
\varepsilon \geq 1$ in the limit of $k\rightarrow 0$, there are no
antisymmetric states at $\varepsilon <1$. The integral power of the
antisymmetric state is%
\begin{equation}
P_{\mathrm{anti}}(k)=\frac{1}{2}U_{\mathrm{anti}}^{2}(k)\left[ \frac{1+\coth
\left( \sqrt{k/2}\right) }{\sqrt{k/2}}-\frac{1}{\sinh ^{2}\left( \sqrt{k/2}%
\right) }\right] .  \label{Nantisymm}
\end{equation}%
This dependence is displayed in Fig. \ref{fig2b} for $\varepsilon =2$. Note
that expression (\ref{Nantisymm}) with all values of $\varepsilon $
satisfies the above-mentioned anti-VK criterion, $dP/dk<0$, which is
necessary for the stability of bound states supported by the SDF
nonlinearity \cite{HS}.


With the variation of $k$ from the largest value, $k_{\max }$, which is
determined by Eq. (\ref{anti}), towards $k\rightarrow 0$, the power (\ref%
{Nantisymm}) monotonously increases from $P_{\mathrm{anti}}(k_{\max })=0$ to
values diverging as
\begin{equation}
P_{\mathrm{anti}}(k)\approx \left( \varepsilon -1\right) /\sqrt{2k}
\label{divergence}
\end{equation}%
at $k\rightarrow 0$. The divergence is explained by the fact that, in the
limit of $k\rightarrow 0$, there is an antisymmetric delocalized state with
divergent power,%
\begin{eqnarray}
U_{\mathrm{anti}}(x;k &=&0)=\sqrt{\varepsilon -1}\mathrm{sgn}x,~\mathrm{at~}%
x>+1/2~\mathrm{or~}x<-1/2,  \notag \\
U_{\mathrm{anti}}(x;k &=&0)=2\sqrt{\varepsilon -1}x,~\mathrm{at~}|x|<1/2.
\label{k = 0}
\end{eqnarray}

The bound state with broken antisymmetry is given by Eqs. (\ref{left})-(\ref%
{middle}) with amplitudes%
\begin{eqnarray}
\left( U_{\mathrm{broken-anti}}\right) _{1}(k) &=&\sqrt{-\frac{E(k)-\sqrt{%
E^{2}(k)-4}}{2S(k)}},  \label{U1anti} \\
\left( U_{\mathrm{broken-anti}}\right) _{2}(k) &=&\sqrt{-\frac{E(k)+\sqrt{%
E^{2}(k)-4}}{2S(k)}}.  \label{U2anti}
\end{eqnarray}%
This solution exists under condition $E<-2$. The substitution of expression (%
\ref{F}) in the latter condition leads to the following existence area for
the solutions with broken antisymmetry:%
\begin{equation}
\varepsilon \geq \left( \varepsilon _{\min }\right) _{\mathrm{broken-anti}%
}\equiv \sqrt{2k}\frac{1+2\exp \left( -\sqrt{2k}\right) }{1-\exp \left( -2%
\sqrt{2k}\right) }>\left( \varepsilon _{\min }\right) _{\mathrm{anti}},
\label{crit-anti}
\end{equation}%
cf. Eq. (\ref{anti}). This area is located above the blue boundary in Fig. %
\ref{fig1}(b). Because Eq. (\ref{crit-anti}) yields $\varepsilon \geq 3/2$
in the limit of $k\rightarrow 0$, there are no states with broken
antisymmetry at $\varepsilon <3/2$.

In agreement with the existence of the delocalized antisymmetric state (\ref%
{k = 0}), there is also a delocalized state with broken antisymmetry,
\textit{viz}.,%
\begin{eqnarray}
U_{\mathrm{broken-anti}}(x;k &=&0)=U_{-},~\mathrm{at~}x<-1/2,  \notag \\
U_{\mathrm{broken-anti}}(x;k &=&0)=\frac{1}{2}\left( U_{+}+U_{-}\right)
+\left( U_{+}+U_{-}\right) x,~\mathrm{at~}|x|<-1/2,  \notag \\
U_{\mathrm{broken-anti}}(x;k &=&0)=U_{+},~\mathrm{at~}x>-1/2,
\label{broken k=0}
\end{eqnarray}%
where%
\begin{equation}
U_{\pm }=\sqrt{\frac{1}{2}\left[ \left( \varepsilon -\frac{1}{2}\right) \pm
\sqrt{\left( \varepsilon -\frac{3}{2}\right) \left( \varepsilon +\frac{1}{2}%
\right) }\right] }.  \label{Upm}
\end{equation}%
The mirror image of this solution is also a delocalized state with broken
antisymmetry. Note that the delocalized antisymmetric state and the one with
the broken antisymmetry exist, according to Eqs. (\ref{k = 0}) and (\ref{Upm}%
), at $\varepsilon >1$ and $\varepsilon >3/2$, respectively, in accordance
with what is said above for the generic solutions of the same types.

For the comparison's sake, it is relevant to mention that Eq. (\ref%
{stationary}) with the SF nonlinearity, $\sigma =+1$, and $\varepsilon <1$
also gives rise to a delocalized antisymmetric state with $k=0$, \textit{viz}%
.,%
\begin{eqnarray}
U_{\mathrm{anti}}^{(\sigma =+1)}(x;k &=&0)=\sqrt{1-\varepsilon }\mathrm{sgn}%
x,~\mathrm{at~}x>+1/2~\mathrm{or~}x<-1/2,  \notag \\
U_{\mathrm{anti}}^{(\sigma =+1)}(x;k &=&0)=2\sqrt{1-\varepsilon }x,~\mathrm{%
at~}|x|<1/2,  \label{k=0 SF}
\end{eqnarray}%
cf. Eq. (\ref{k = 0}). However, as well as all antisymmetric solutions of
Eq. (\ref{psi}) with the SF nonlinearity, this solution is unstable (against
modulational perturbations, cf. Rev. \cite{Azbel}), and Eq. (\ref{stationary}%
) with $\sigma =+1$ does not produce solutions with unbroken antisymmetry.

\section*{Numerical solutions}

Numerical investigation of Eqs. (\ref{psi}) and (\ref{stationary}) with the $%
\delta $-functions approximated as per Eq. (\ref{tilde}) is relevant for
checking the analytical results reported above. Because the width of the
linear and nonlinear potential wells in real systems is finite, the
numerical results are also relevant for the verification of the relevance of
the analytical predictions, which are obtained above with the use of the
ideal $\delta $-functions.

\subsection*{The self-focusing nonlinearity}

\begin{figure}[tbph]
\includegraphics[scale=0.35]{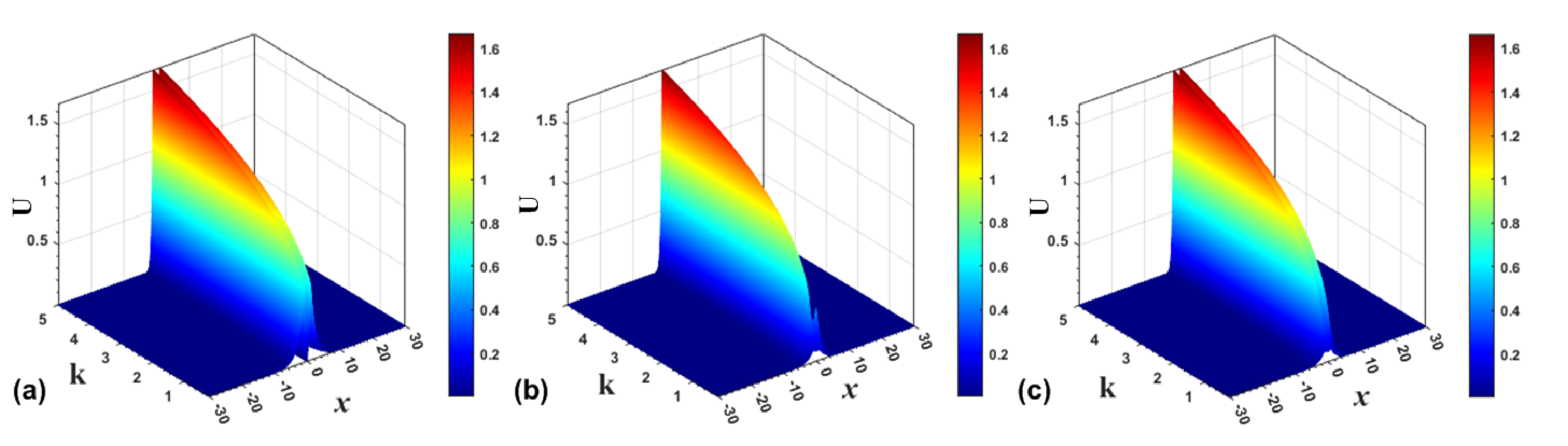}
\caption{The evolution of the shapes of antisymmetric (a), asymmetric (b),
and symmetric (c) numerically produced solutions of Eq. (\protect\ref%
{stationary}) with $\protect\sigma =+1$ and $\protect\varepsilon =0.5$,
following the increase of $k$.}
\label{fig4}
\end{figure}

Numerically found examples of bound states of the symmetric and
antisymmetric types, as well as ones with broken symmetry and antisymmetry,
in the cases of the SF and SDF signs of the nonlinearity, are displayed
above in Fig. \ref{fig0}. In a systematic way, the evolution of the
antisymmetric, asymmetric, and symmetric solitons produced by Eq. (\ref%
{stationary}) with $\sigma =+1$, following the increase of propagation
constant $k$, is summarized in Fig. \ref{fig4}.

\begin{figure}[tbph]
\includegraphics[scale=0.25]{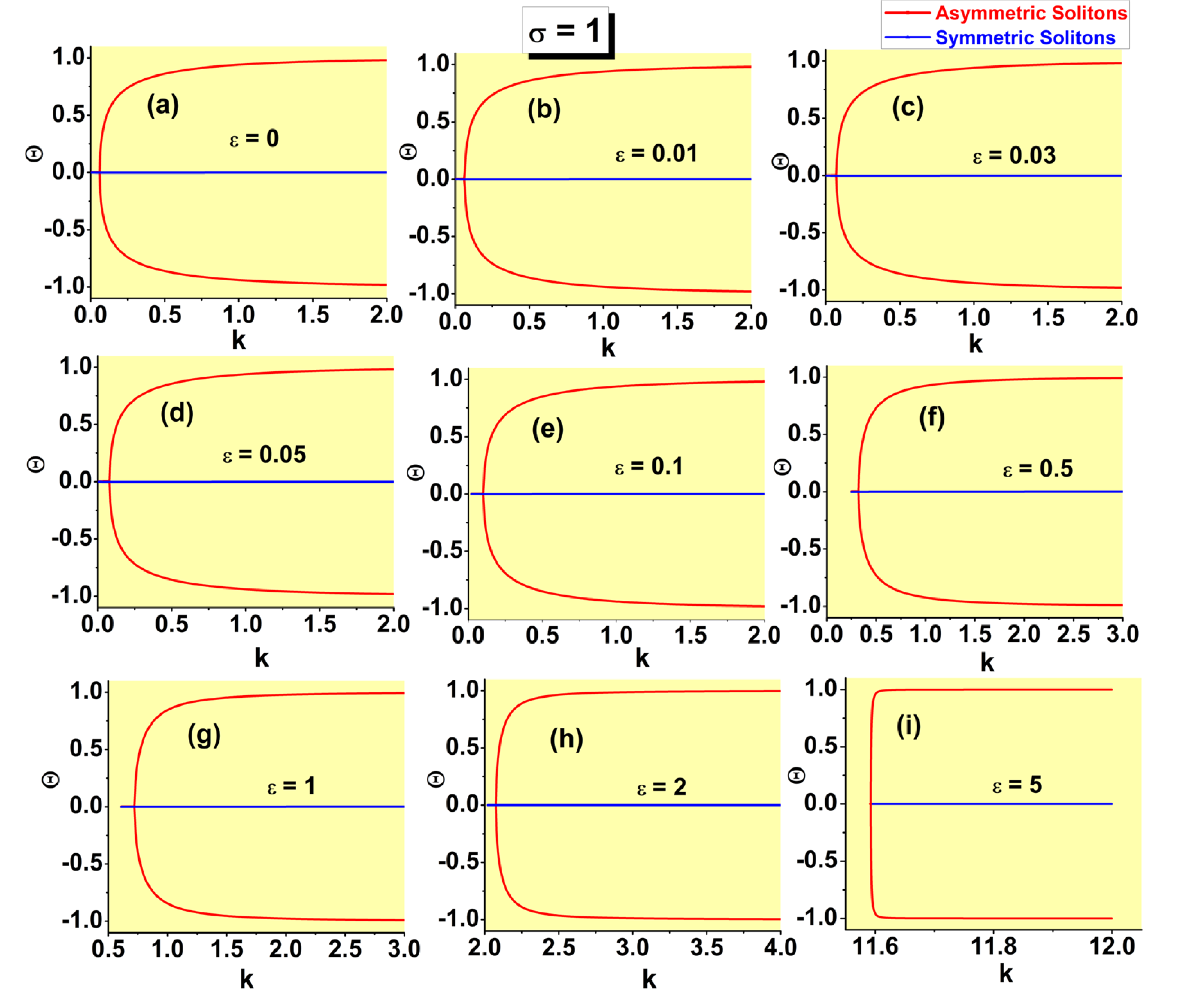}
\caption{The asymmetry parameter (\protect\ref{Theta}) for the numerically
produced solutions of Eq. (\protect\ref{stationary}) with $\protect\sigma =+1
$ vs. propagation constant $k$ at the same values of $\protect\varepsilon $
which are presented in Fig. \protect\ref{fig3}.}
\label{fig5}
\end{figure}

The most essential results in the form of the SSB diagrams for the SF\
model, which corroborate the basic analytically predicted property of the
model, \textit{viz}., the switch of the character of the symmetry-breaking
phase transition from the first to second kind (in other words, the switch
from the subcritical SSB bifurcation to the supercritical one) at the
threshold point (\ref{thr}), are also demonstrated above in Fig. \ref{fig3}.
In addition to that, it is relevant to plot the bifurcation diagrams in the
plane of $k$ and asymmetry parameter $\Theta $. These are presented in Fig. %
\ref{fig5}, for the same set of values of $\varepsilon $ as in Fig. \ref%
{fig3}. The branch of the symmetric states commences at $k=\left( k_{\min
}\right) _{\mathrm{symm}}$ [see Eq. (\ref{kmin})], while the value of $%
k(\varepsilon )$ at the SSB bifurcation point is determined by Eq. (\ref%
{crit}).

\begin{figure}[tbph]
\includegraphics[scale=0.25]{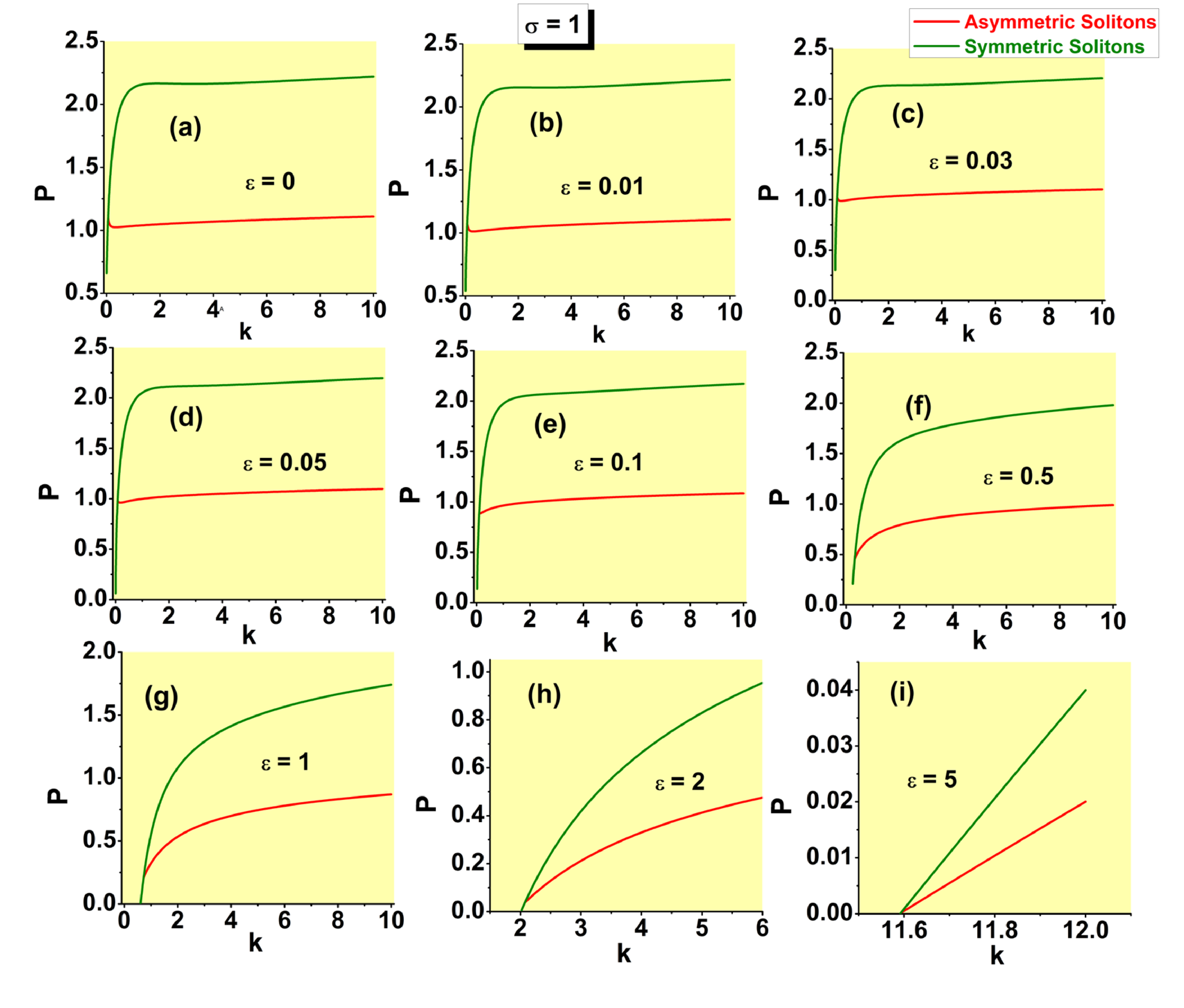}
\caption{The integral power of the symmetric and asymmetric bound states in
the case of the SF nonlinearity ($\protect\sigma =+1$) vs. the propagation
constant for the same values of $\protect\varepsilon $ as in Figs. \protect
\ref{fig3} and \protect\ref{fig5}.}
\label{fig6}
\end{figure}

\begin{figure}[tbph]
\includegraphics[scale=0.25]{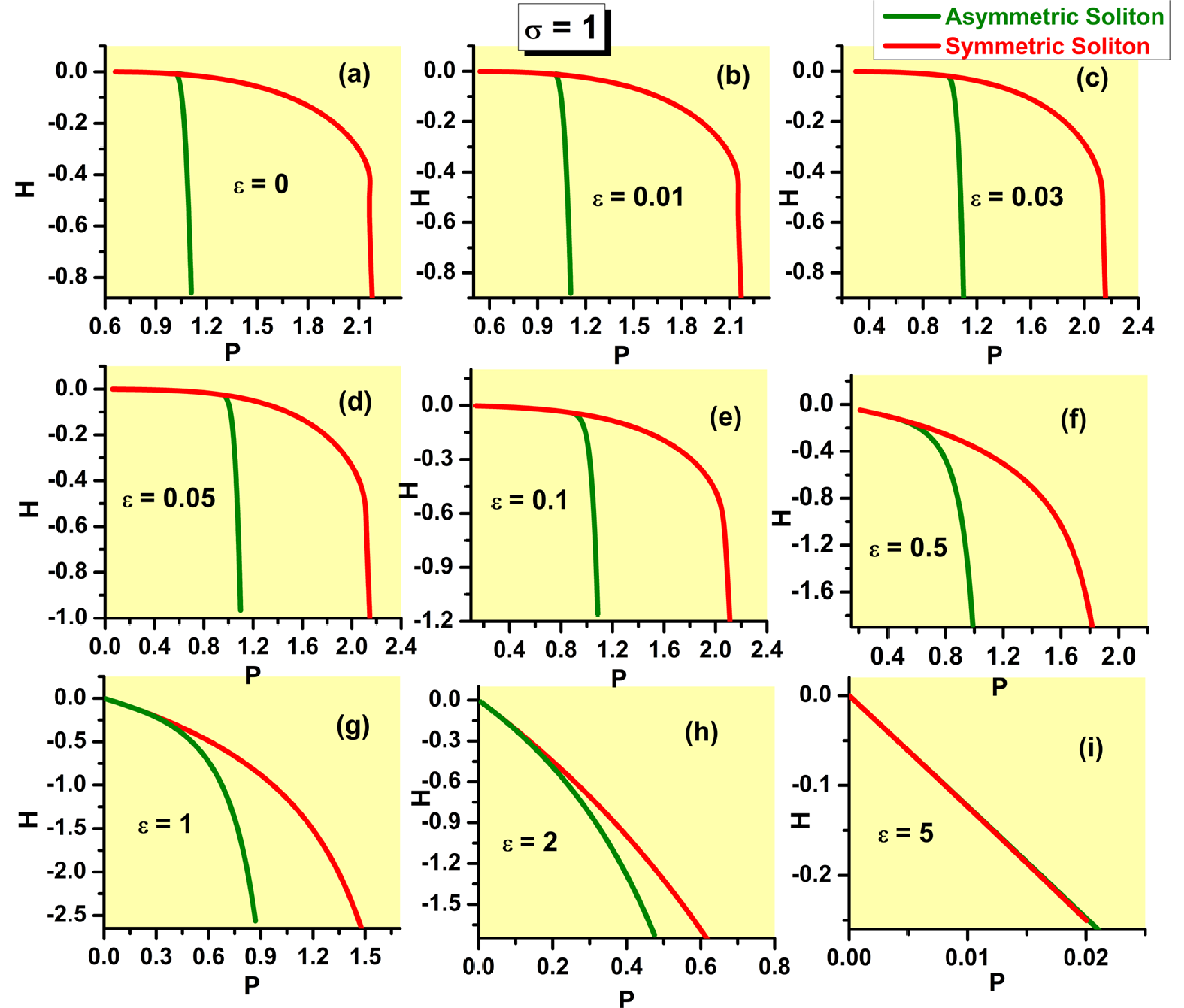}
\caption{The Hamiltonian of the symmetric and asymmetric bound states,
calculated as per Eq. (\protect\ref{Hamiltonian}) with $\protect\sigma =+1$,
vs. the integral power, $P$, for the same values of $\protect\varepsilon $
as in Figs. \protect\ref{fig3}, \protect\ref{fig5}, and \protect\ref{fig6}.}
\label{fig7}
\end{figure}

Further, the families of symmetric and asymmetric solitons are
characterized, as physical states, by the respective dependences $P(k)$ and $%
H(P)$, where $H$ is the Hamiltonian defined by Eq. (\ref{Hamiltonian}).
These dependences are displayed, respectively, in Figs. \ref{fig6} and \ref%
{fig7}. In the former figure, the branches of the symmetric states commence
at $k=\left( k_{\min }\right) _{\mathrm{symm}}$, see Eq. (\ref{kmin}). In
panels (a-f) of Fig. \ref{fig6}, $P$ varies between limit values $0$ and $2$
along the symmetric branches, and between $P_{\mathrm{bif}}$ [see Eq. (\ref%
{Pbif})] and $P=1$ along the the asymmetric ones [in panels (g-i), the
variation range of $P$ is truncated; it is also partly cut in Fig. \ref{fig7}%
(f)]. In Fig. \ref{fig7}(i), dependences $H(P)$ for the symmetric and
asymmetric states are indistinguishable. Note also that, in latter case, the
value $P_{\mathrm{bif}}$ at the SSB point is extremely small, in agreement
with Eq. (\ref{exp}). On the other hand, coordinates of the SSB points in
Figs. \ref{fig6}(a) and \ref{fig7}(a) are correctly predicted by Eqs. (\ref%
{eps=0}) and (\ref{Pbif-max}).

In the range where the asymmetric states exist, they realizes the minimum of
$H$, i.e., the system's GS. A specific feature of the system is that it has
no true GS at larger values of $P$, where only the \emph{unstable} symmetric
states exist, and there are no stationary states whatsoever at $P>2$.

\subsection*{The self-defocusing nonlinearity}

\begin{figure}[tbph]
\includegraphics[scale=0.25]{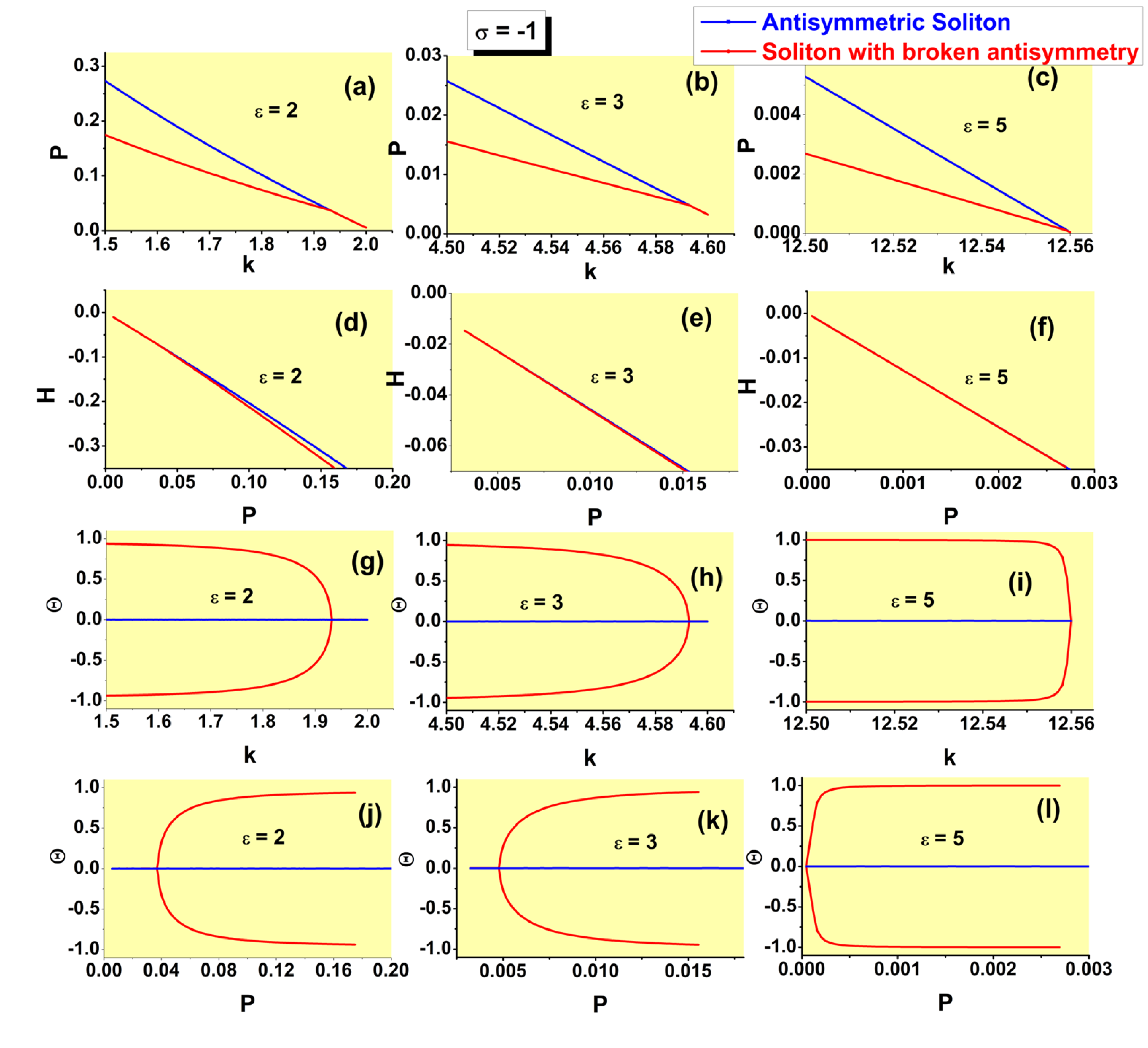}
\caption{Dependences $P(k)$ (a-c), $H(P)$ (d-f), $\Theta (k)$ (g-i), and $%
\protect\theta (P)$ (j-l) for the families of antisymmetric solitons and
those with broken antisymmetry, generated by the numerical solution of Eq. (%
\protect\ref{stationary}) with $\protect\sigma =-1$, three different values
of $\protect\varepsilon $, which are indicated in the panels, and the $%
\protect\delta $-function approximated by expression (\protect\ref{tilde}).
Counterparts of these dependences in the system with $\protect\sigma =+1$
are displayed above in Figs. \protect\ref{fig6}, \protect\ref{fig7}, \protect
\ref{fig5}, and \protect\ref{fig3}, respectively.}
\label{fig8}
\end{figure}

Dependences $P(k)$, $H(P)$, $\Theta (k)$, and $\theta (P)$ for the families
of antisymmetric solitons and those with broken antisymmetry, as produced by
the numerical solution of Eq. (\ref{stationary}) with $\sigma =-1$, are
collected, severally, in panels (a-c), (d-f), (g-i), and (j-l) of Fig. \ref%
{fig8}, for three different values of the strength of the linear $\delta $%
-function potential, \textit{viz}., $\varepsilon =2,3,$ and $5$. These sets
of plots are counterparts of those for the model with $\sigma =+1$ which are
displayed above in Figs. \ref{fig6}, \ref{fig7}, \ref{fig5}, and \ref{fig3},
respectively. In particular, the $P(k)$ curves and SSB points on all the
curves plotted in Fig. \ref{fig8} are correctly predicted by Eqs. (\ref%
{Nantisymm}) and (\ref{anti}), respectively.

An obvious difference from the case of the SF nonlinearity is that the
bifurcation of the spontaneous breaking of antisymmetry in the SDF case is
always supercritical, as seen in Figs. \ref{fig8}(j-l). In other words, the
model with the SDF nonlinearity always gives rise to the
antisymmetry-breaking phase transition of the second kind. It is also worthy
to note that the soliton branches with both unbroken and broken antisymmetry
always satisfy the above-mentioned anti-VK criterion, $dP/dk<0$, which is
the necessary (but not sufficient) condition for their stability.

\begin{figure}[tbph]
\includegraphics[scale=0.35]{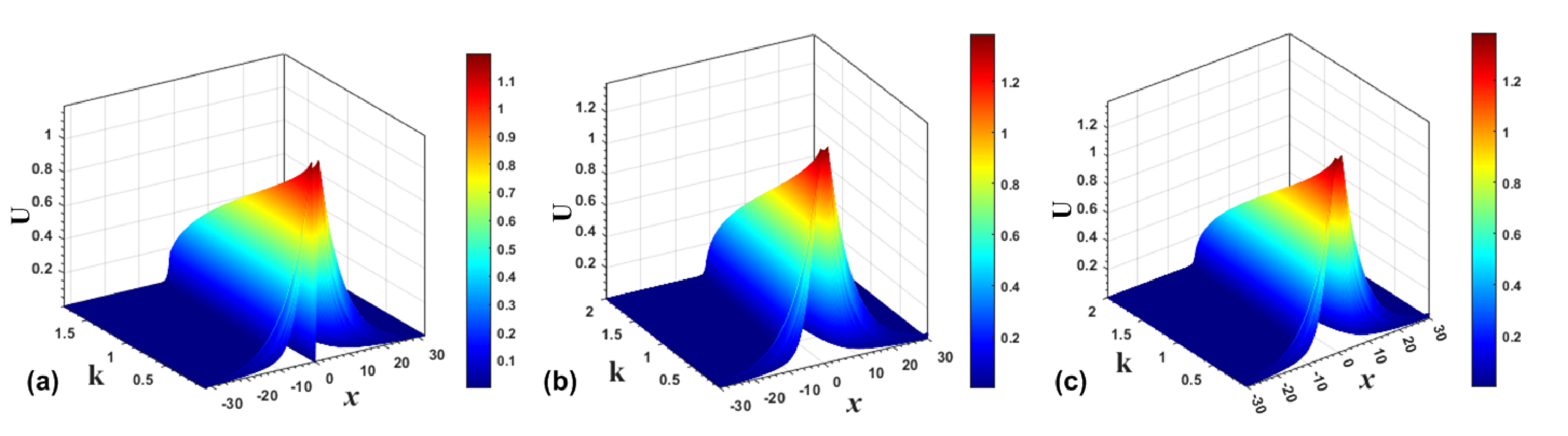}
\caption{The evolution of shapes of the antisymmetric (a),
broken-antisymmetry (b), and symmetric (c) numerically found solutions of
Eq. (\protect\ref{stationary}) with $\protect\sigma =-1$ and $\protect%
\varepsilon =2$, following the increase of $k$.}
\label{fig9}
\end{figure}

Finally, the evolution of the antisymmetric, broken-antisymmetry, and
symmetric bound states produced by Eq. (\ref{stationary}) with $\sigma =-1$
and $\varepsilon =2$, following the increase of propagation constant $k$, is
summarized in Fig. \ref{fig9}. Note that, in agreement with the analytical
solutions, the evolution is opposite to that in the model with the SF
nonlinearity ($\sigma =+1$), which is displayed above in Fig. \ref{fig4}.
Namely, the amplitude and integral power of the solitons increase/decrease
with the growth of $k$ in the SF/SDF system.

\subsection*{The evolution of unstable bound states}

It is relevant to test the expected (in)stability of symmetric and
antisymmetric bound states, as well as ones with broken symmetry and
antisymmetry, in direct simulations of Eq. (\ref{psi}) with the ideal $%
\delta $-function replaced by its regularized version (\ref{tilde}), for
both the SF\ and SDF signs of the nonlinearity, i.e., $\sigma =+1$ and $-1$.

\begin{figure}[tbph]
\includegraphics[scale=0.5]{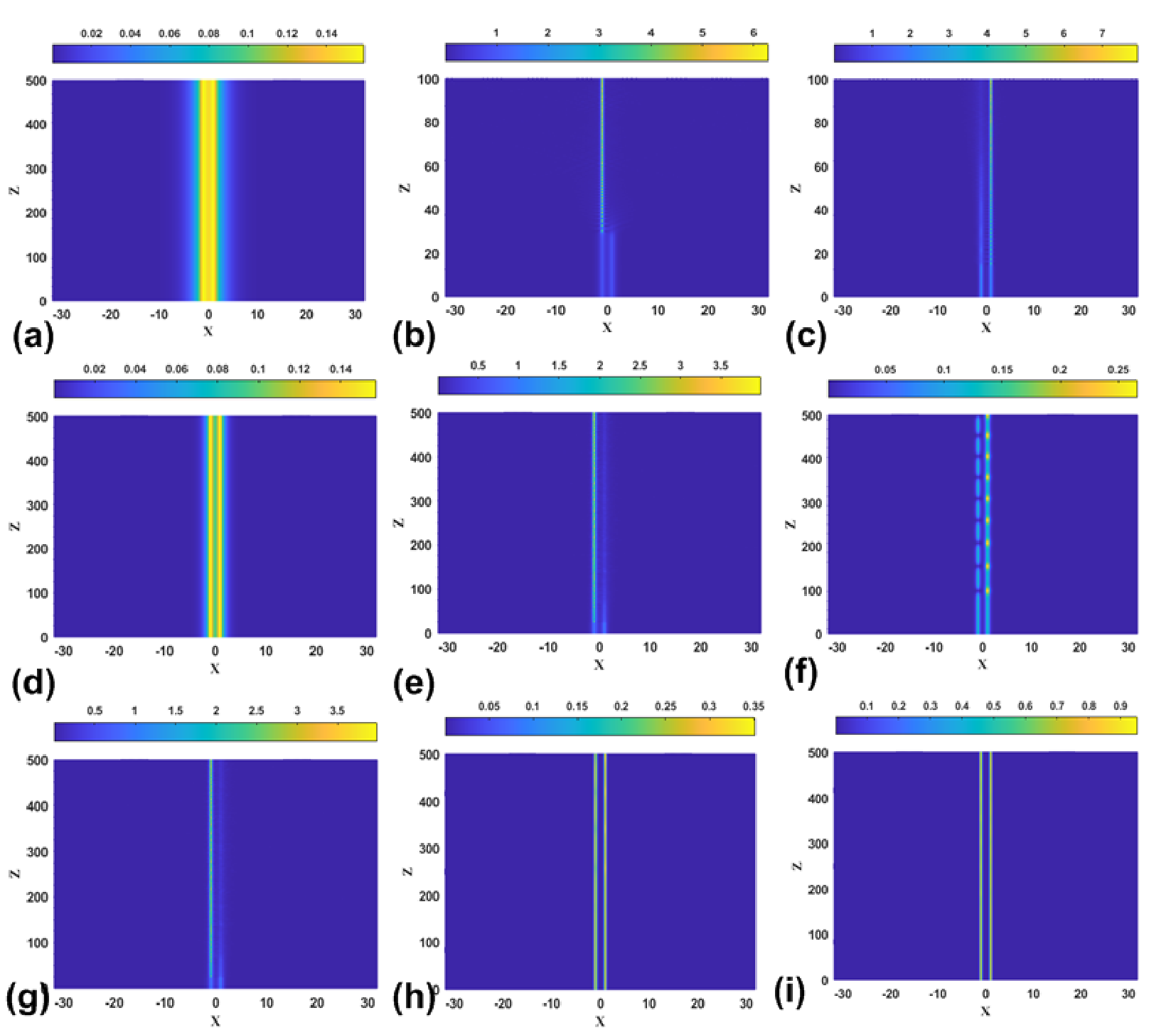}
\caption{The evolution of stable and unstable symmetric bound states in the
model with the SF nonlinearity, as produced by simulations of Eq. (\protect
\ref{psi}) with $\protect\sigma =+1$ and parameters $(\protect\varepsilon %
,k)=(0.05,0.05)$ (a), $(0.05,0.1)$ (b), $(0.5,2)$ (c), $(0.5,0.3)$ (d), $%
(0.5,1)$ (e), $(2,2)$ (f), $(2,2.8)$ (g), $(5,10)$ (h), and $(5,11.7)$ (i).
The panels plot values of $\left\vert \protect\psi (x,z)\right\vert $ by
means of the color code.}
\label{fig10}
\end{figure}

First, Fig. \ref{fig10} collects typical examples which demonstrate the
perturbed evolution of stable [panels (a,d,f,h,i)] and unstable [panels
(b,c,e,g)] symmetric bound states in the model with the SF nonlinearity.
These results are compatible with the prediction of the stability area for
the symmetric states, in the form of the stripe between the lower and upper
boundaries in Fig. \ref{fig1}(a). It is observed that, naturally, all the
unstable symmetric states demonstrate manifestations of the SSB instability,
leading to spontaneous formation of asymmetric states. In some cases, such
as the one displayed in panel \ref{fig1}(f), the unstable symmetric state,
which is located close to the instability boundary, features conspicuous
persistent oscillations, while in other cases the stronger instability
creates nearly stationary modes with strong asymmetry.
{\LARGE \ }

\begin{figure}[tbph]
\includegraphics[scale=0.5]{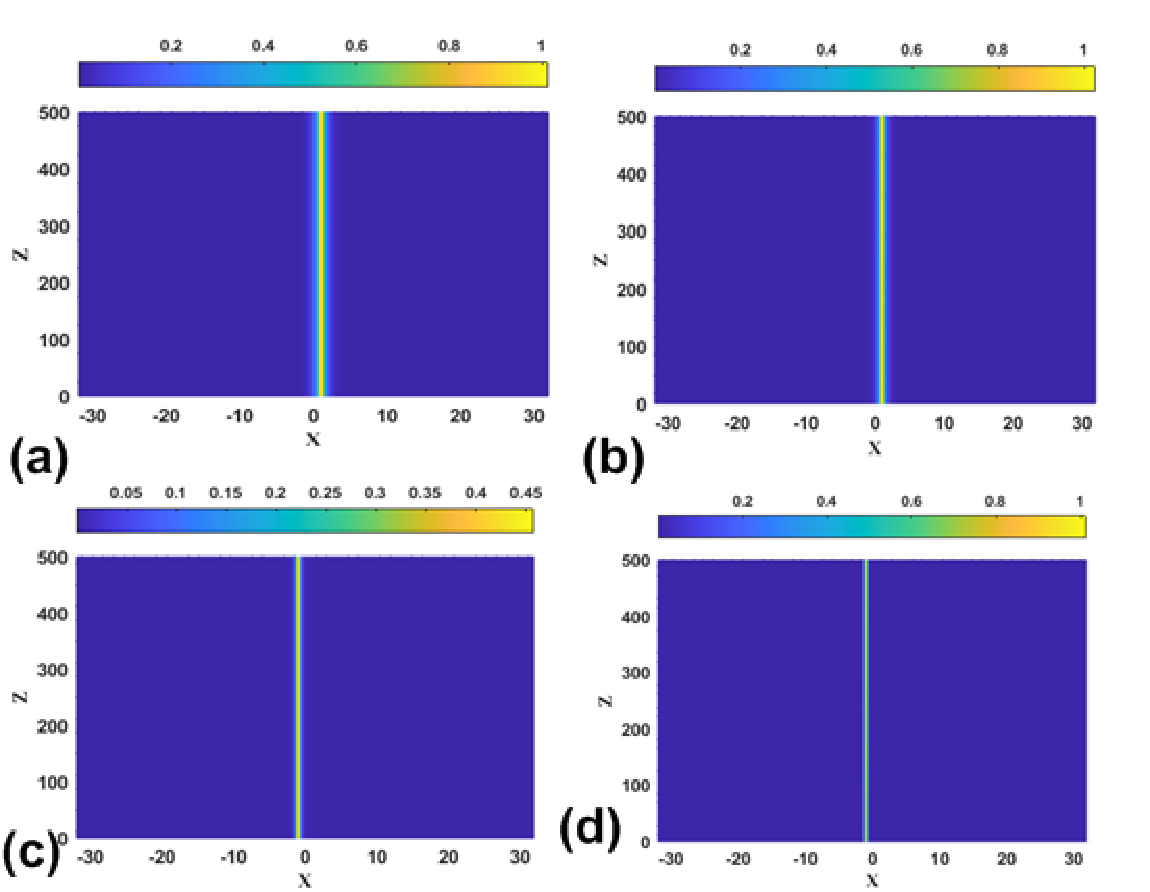}
\caption{The evolution of stable asymmetric bound states in the model with
the SF sign of the nonlinearity ($\protect\sigma =+1$), as produced by
simulations of Eq. (\protect\ref{psi}) with $\protect\sigma =+1$ and
parameters $(\protect\varepsilon ,k)=(0.05,0.5)$ (a), $(0.5,1)$ (b), $(2,2.5)
$ (c), $(5,12)$ (d).}
\label{fig11}
\end{figure}

Another expected result corroborated by the direct simulations of the
perturbed evolution is that nearly all the asymmetric solitons are stable in
the case of the SF nonlinearity, as shown in Fig. \ref{fig11} for strongly
asymmetric solutions. Unstable are asymmetric solitons belonging to the
backward-going (lower) branch in Figs. \ref{fig3}(a-d). In fact, they exist
only in a very narrow parameter region, and the development of the
instability pulls them towards a stable counterpart existing at the same
value of $P$ (not shown here in detail, as this is a known feature of the
subcritical SSB bifurcation).

In addition to the above results, direct simulations displayed in Fig. \ref%
{fig12} confirm the expected instability of all antisymmetric bound states
in the case of the SF nonlinearity. In the cases shown in panels (e) and (f)
of the figure, the instability is barely visible as the interaction between
two power peaks of the antisymmetric modes is very weak.
\begin{figure}[tbph]
\includegraphics[scale=0.5]{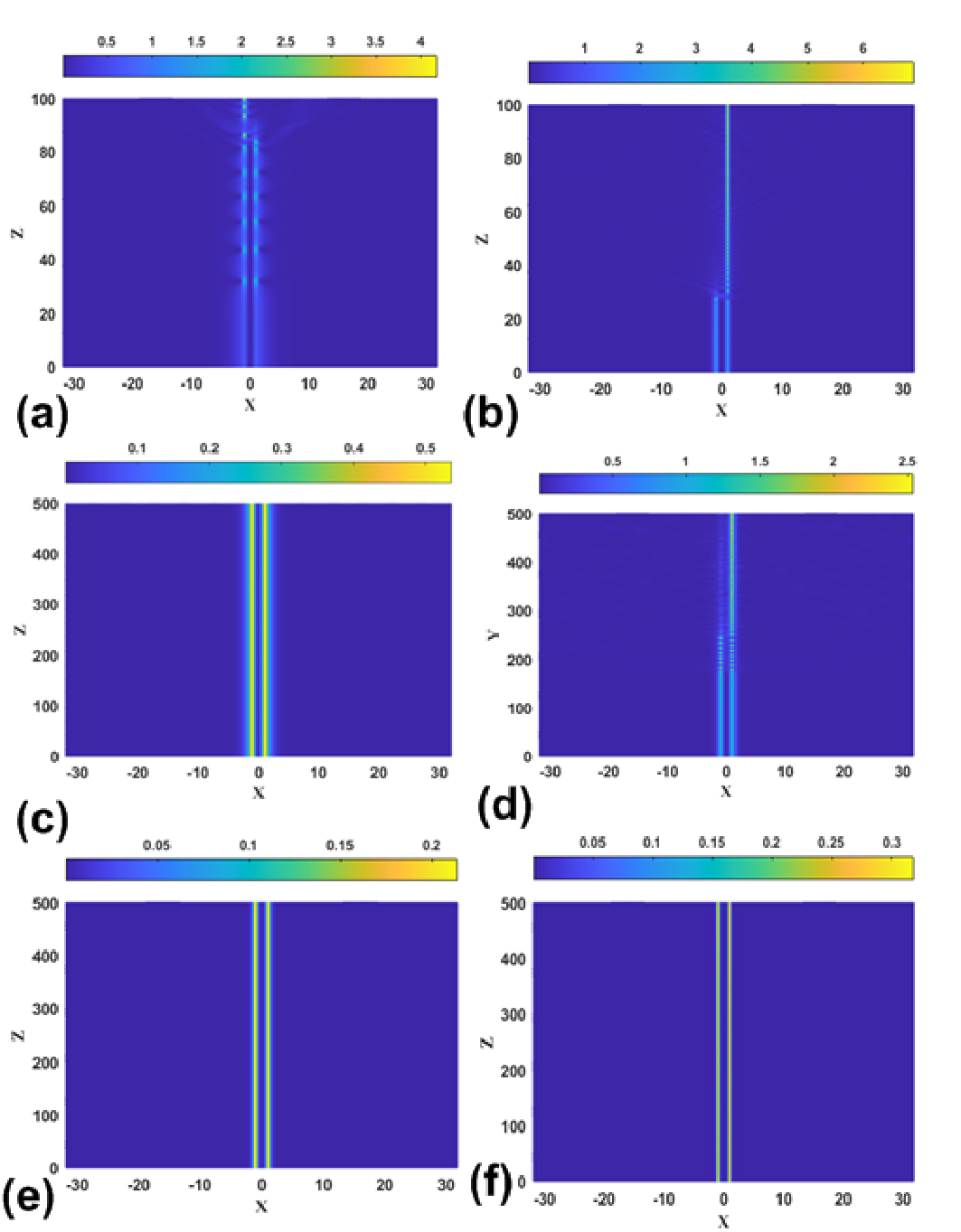}
\caption{The evolution of unstable antisymmetric bound states in the model
with the SF nonlinearity, as produced by simulations of Eq. (\protect\ref%
{psi}) with $\protect\sigma =+1$ and parameters $(\protect\varepsilon %
,k)=(0.05,0.05)$ (a), $(0.05,2)$ (b), $(0.5,0.3)$ (c), $(0.5,1)$ (d), $(2,2)$
(e), $(5,10)$ (f).}
\label{fig12}
\end{figure}

\begin{figure}[tbph]
\includegraphics[scale=0.5]{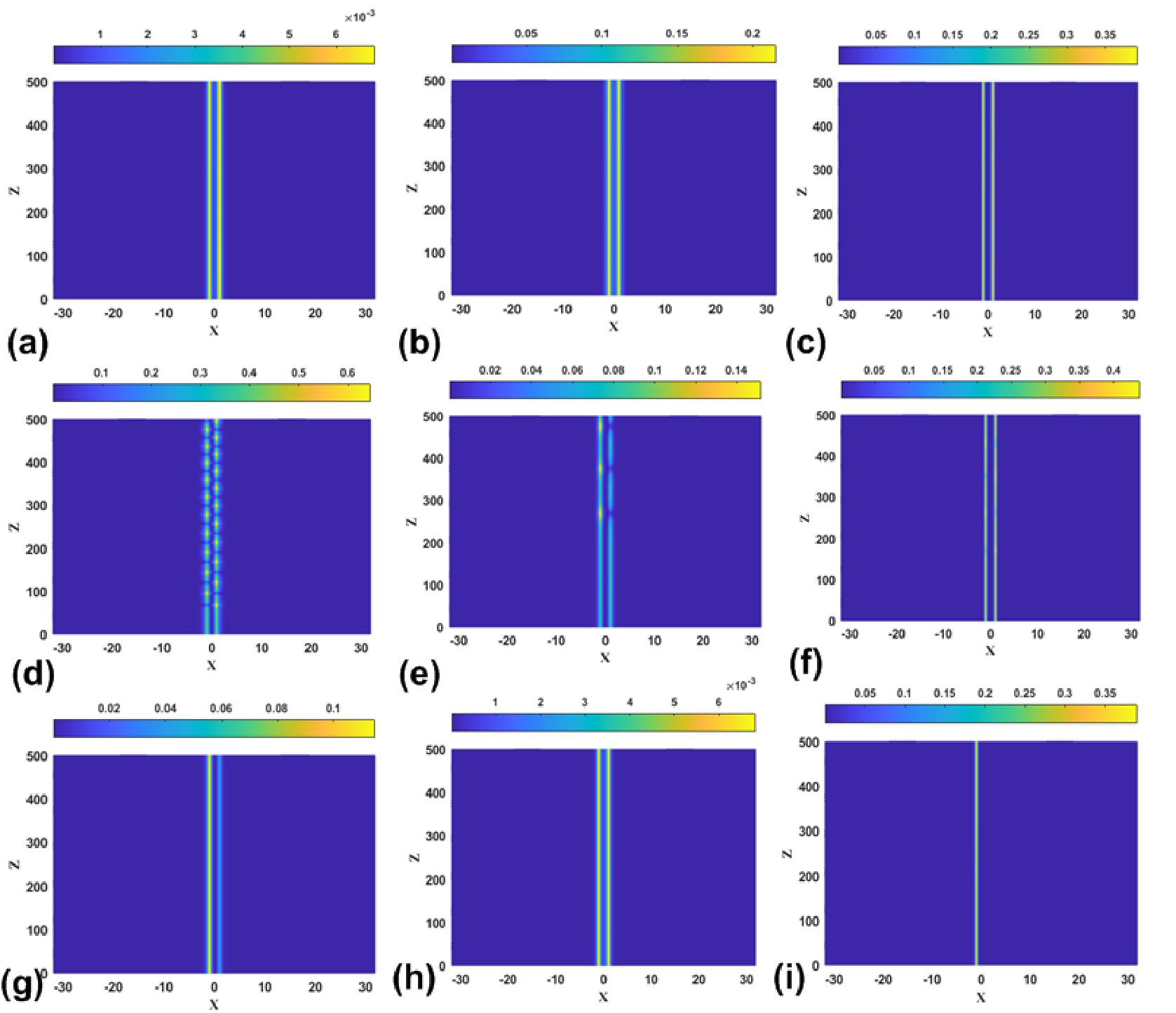}
\caption{(a), (b) and (c): The evolution of stable symmetric bound states in
the model with the SDF sign of the nonlinearity, as produced by simulations
of Eq. (\protect\ref{psi}) with $\protect\sigma =-1$ and parameters $(%
\protect\varepsilon ,k)=(2,1.5)$, $(2,2)$ and $(5,8)$, respectively. (d,e)
and (f:): The evolution of moderately and weakly unstable antisymmetric
bound states for $(\protect\varepsilon ,k)=(2,1)$, $(2,1.5)$, and $(5,8)$,
respectively. (g), (h) and (i): The evolution of stable bound states with
moderately, weakly, and strongly broken antisymmetry, for $(\protect%
\varepsilon ,k)=(2,1.5)$, $(2.8,1.8)$ and $(5,8)$, respectively.}
\label{fig13}
\end{figure}

Lastly, characteristic examples of the perturbed evolution of the bound
states of the symmetric, antisymmetric, and broken-antisymmetry types in the
model with the SDF nonlinearity are collected in Fig. \ref{fig13}. In
particular,\ in agreement with the above predictions all symmetric states
are stable in this case, as shown in Figs. \ref{fig13}(a-c). Further, panels
(d,e) and (f) of Fig. \ref{fig13} present, respectively, examples of
moderately and weakly unstable antisymmetric states, in agreement with the
boundaries plotted in Fig. \ref{fig1}(b). It is seen that the instability
leads to spontaneous replacement of the corresponding states by oscillatory
ones with broken antisymmetry. Finally, panels (g-i) demonstrate stability
of the stationary states with weakly or strongly broken antisymmetry, also
in agreement with the boundary plotted in Fig. \ref{fig1}(b).

\section*{Discussion}

Starting from the two-dimensional Ising lattice \cite{Ising,LLstat}, exactly
solvable models serve as benchmarks for studies of phase transitions in
diverse physical settings \cite{Baxter}-\cite{Rosten}. Transitions from a
paramagnetic phase to a ferromagnetic one in spin systems, and similar
transitions in many other media are intrinsically related to spontaneous
breaking of the symmetry of the underlying setting. It is well known that
phase transitions are classified as ones of the first and second kinds.
Hysteresis and bistability, in the form of the coexistence of the GS (ground
state) with a metastable overcooled or overheated phase, are possible in the
former case.

Similar phenomenology is exhibited by nonlinear dynamical systems, in the
form of bifurcations, i.e., transitions between different stable states of
the system, caused by variation of the system's control parameter(s) \cite%
{bif}. Counterparts of the phase transitions of the first and second kinds
are identified as bifurcations of the subcritical and supercritical types in
dynamical systems. The subcritical bifurcation creates stable states prior
to the destabilization of the symmetric one, thus the bifurcation of this
type admits the bistability and hysteresis, like phase transitions of the
first kind.

In most cases, phase transitions in statistical physics, as well as
bifurcations in dynamical systems, are studied between spatially uniform
states. On the other hand, transitions between spatially localized
(self-trapped) modes, such as solitons, are possible too. The analysis of
the latter topic may benefit from the consideration of models admitting
exact solutions for symmetry-breaking transitions in self-trapped states.
However, finding solvable models is a challenging task, because basic
integrable models that give rise to solitons, such as the one-dimensional
NLSE, do not admit intrinsic transitions in the solitons.

The objective of the present work is to introduce a solvable nonlinear model
with the DWP (double-well potential) which makes it possible to produce
exact solutions for localized states with full and broken symmetries, that
are linked by symmetry-breaking transitions of both first and second kinds.
In other words, the states with unbroken and broken symmetries may be linked
by bifurcations of the sub- and supercritical types. The integrability of
the present model is possible due to the fact that the nonlinearity is
represented by the symmetric set of two $\delta $-functions. A prototype of
this model was introduced previously in Ref. \cite{Thawatchai}, but it had
produced a very limited result, \textit{viz}., the SSB
(spontaneous-symmetry-breaking) bifurcation of the extreme subcritical form.
That bifurcation gave rise to completely unstable asymmetric states,
represented by backward-going solution branches which never turned forward.
In the present work, we have introduced the solvable DWP model including
both nonlinear and linear potentials, which are based on the symmetric pair
of $\delta $-functions. The respective nonlinear potential is considered
with both the SF (self-focusing) and SDF\ (self-defocusing) signs.

The analytical solutions, confirmed by their numerically found counterparts
(which were produced replacing the ideal $\delta $-functions by the narrow
Gaussians), give rise to the full set of symmetric and asymmetric states in
the model with the SF nonlinearity, as well as the full set of symmetric and
antisymmetric states, along with ones with broken antisymmetry, in the SDF
model. In the case of the SF nonlinearity, the most important aspect of the
analytical solution is the explicitly found switch from the
symmetry-breaking phase transition of the first kind into one of the second
kind, or, in other words, the switch of the subcritical bifurcation into the
supercritical one. The switch takes place with the increase of strength $%
\varepsilon $ of the linear part of the DWP potential based on the symmetric
pair of $\delta $-functions. Starting from the above-mentioned extreme
subcritical bifurcation at $\varepsilon =0$, the switch is found
analytically to occur at the point given by Eqs. (\ref{eq}) and (\ref{thr}),
which is corroborated by the numerical findings. To the best of our
knowledge, no previously studied model made it possible to predict the
change of a symmetry-breaking phase transition between the first and second
kinds (or the change of the sub/supercritical character of the SSB
bifurcation) in an analytical form.

The analytical solution is also reported here for the model with the SDF
nonlinearity, where the situation is simpler: the GS is always represented
by the completely stable symmetric localized state, while the
antisymmetry-breaking phase transition of the second kind (i.e., the
supercritical bifurcation) destabilizes the lowest excited state (a
spatially odd stationary one) at the critical point given by Eq. (\ref%
{crit-anti}). These analytical results for the SDF model are confirmed by
the numerical solution too.

The solvable models elaborated in the present work suggest possibilities for
analytical studies of SSB phase transitions and bifurcations in more complex
settings. In particular, it may be interesting to address a two-component
system with the combined linear-nonlinear DWP potential. A degenerate form
of the two-component system, with the nonlinear-only SF potential, based on
the symmetric pair of $\delta $-functions, was introduced in Ref. \cite%
{Yasha}. In that model, the SF\ nonlinearity includes self-interaction in
each component and cross-interaction between the components. Note that the
two-component SF model, unlike the single-component one, admits an
antisymmetry-breaking phase transition in spatially odd localized states,
and it also opens the way to the consideration of the SSB transition in a
state which combines spatially symmetric and antisymmetric components.

Another new possibility is offered by a model with three equidistant $\delta
$-functions set on a circle, unlike the infinite one-dimensional domain
considered in the present work (the circle with the purely nonlinear SDF
potential, represented by a symmetric pair of $\delta $-functions set at
diametrically opposite points, was addressed in Ref. \cite{Han-Pu}, in which
case it did not give rise to SSB transitions, the respective GS being always
symmetric). Various setups with a triangle of potential wells embedded in a
nonlinear medium were studied for BEC\ \cite{triple-BEC1,triple-BEC2} and
multicore optical fibers \cite{SotoCrespo}-\cite{Porsezian}. In the circular
setting with three $\delta $-function wells, one can construct exact
solutions carrying vorticity and address feasible SSB transitions in them.

On the other hand, as a step towards the consideration of two-dimensional
models, where full solvability is not plausible, one can introduce a set of
two parallel linearly-coupled one-dimensional lines, each bearing a DWP
represented by the symmetric pair of the $\delta $-functions. In all these
extensions, analytical solutions will take an essentially more cumbersome
form than the one addressed in the present work, but the analysis may still
be possible.

\section*{Data availability}

The data that support the findings of this study are available from the
corresponding author upon reasonable request

\section*{Acknowledgements}

The work of S.K. and B.A.M. is supported, in part, by the Israel Science
Foundation through grant No. 1695/22.

\section*{Author contributions}

B.A.M.: formulation of the model, analytical and numerical calculations,
drafting the manuscript; S.K.: formulation of the model, development of
numerical codes, numerical calculations, and drafting the manuscript; P.L.
and L.Z.: development of numerical codes, numerical calculations, and
drafting the manuscript; J.H.: formulation of the model, numerical
calculations, and drafting the manuscript.

\section*{Competing interests}

The authors declare no competing interests.

\end{document}